\documentclass[useAMS,usenatbib]{mn2e}
\pdfoutput=1

\usepackage{natbib,hyperref,amssymb}
\usepackage{epsfig,graphicx}
\usepackage{color}

\title[ML for Discovery in Synoptic Imaging]{Using Machine Learning for Discovery in Synoptic Survey Imaging} 
\author[H. Brink et al.]
       {\parbox[]{6.0in}
	{Henrik Brink$^{1}$\thanks{E-mail: brink@berkeley.edu}, Joseph W. Richards$^{1,2}$,
Dovi Poznanski$^{3}$, Joshua S. Bloom$^{1}$, John Rice$^{2}$, Sahand Negahban$^{4}$,
Martin Wainwright$^{2,4}$\\
        \footnotesize
$^{1}$Department of Astronomy, University of California Berkeley\\
$^{2}$Department of Statistics, University of California Berkeley\\
$^{3}$School of Physics and Astronomy, Tel Aviv University\\
$^{4}$Department of Electrical Engineering and Computer Sciences, University of California Berkeley}}

\begin{document}


\pagerange{\pageref{firstpage}--\pageref{lastpage}} \pubyear{2012}

\maketitle
\label{firstpage}

\begin{abstract}
Modern time-domain surveys continuously monitor large swaths of the sky to look for astronomical variability.  Astrophysical discovery in such data sets is complicated by the fact that detections of {\it real} transient and variable sources are highly outnumbered by {\it bogus} detections caused by imperfect subtractions, atmospheric effects and detector artefacts. In this work we present a machine learning (ML) framework for discovery of variability in time-domain imaging surveys.  Our ML methods provide probabilistic statements, in near real time, about the degree to which each newly observed source is astrophysically relevant source of variable brightness.  We provide details about each of the analysis steps involved, including compilation of the training and testing sets, construction of descriptive image-based and contextual features, and optimization of the feature subset and model tuning parameters.   Using a validation set of nearly 30,000 objects from the Palomar Transient Factory, we demonstrate a missed detection rate of at most $7.7\%$ at our chosen false-positive rate of $1\%$ for an optimized ML classifier of 23 features, selected to avoid feature correlation and over-fitting from an initial library of 42 attributes. Importantly, we show that our classification methodology is insensitive to mis-labelled training data up to a contamination of nearly $10\%$, making it easier to compile sufficient training sets for accurate performance in future surveys. This ML framework, if so adopted, should enable the maximization of scientific gain from future synoptic survey and enable fast follow-up decisions on the vast amounts of streaming data produced by such experiments. 

\end{abstract}


\section{Introduction}

Synoptic surveys have begun to generate enough imaging data on a nightly basis to significantly tax the ability of humans to inspect each image and search for new events. Automating aspects of reduction and discovery processes is, of course, crucial. But in any imaging survey, large fractions of apparent variability will be not be astrophysically relevant due to random fluctuations, noise, systematic errors in the analysis process, and near-field objects (such as satellite streaks). The task of any automation approach is to effectively separate this wide range of spurious detections from the real---and possibly scientifically valuable---events. The sky is teeming with both transients (such as supernovae, quasars, microlensing) and a host of variable star classes.  With ever growing data volumes, it is becoming clear that we \emph{must} employ automated methods to find such needles in this astronomical haystack. In doing so, we necessarily must supplant the traditional role of humans not just in data analysis but in decision making (e.g., see \citealt{bloom10}). In this paper we detail the development of an automated Machine Learning (ML) framework for this task and use a relevant current survey as a test bed: The Palomar Transient Factory (PTF).

PTF is a synoptic survey imaging with the Palomar 48'' telescope mounted with a refurbished 12-CCD camera, formerly the CFH12K on the Canada France Hawaii Telescope. For details on the survey instruments and strategies see \citet{rau09} and \citet{law09}. The nightly imaging data from PTF is analyzed in real time by a pipeline running at the National Energy Research Scientific Computing Center (NERSC), where the \emph{new} images are aligned to a deep \emph{reference} image of the static sky and the two are subtracted to create a \emph{subtraction} image. Initial candidate detection on the subtraction image is obtained using {\tt SEXTRACTOR} (\citealt{bertin96}; for more details on the pipeline see \citealt{nugent11}). Typically, thousands of variable or transient candidates are detected on each subtraction image, the vast majority of which are subtraction artefacts, which can occur for a plethora of reasons, including  improperly reduced new images, edge effects on the reference or new image, misalignment of the images, improper flux scalings, incorrect PSF convolution, CCD array defects, and cosmic rays. Since larger upcoming surveys, such as the Large Synoptic Survey Telescope (LSST, \citealt{ivezic2008}), will work in much the same way, the PTF survey provides a relevant environment for developing the real-time analysis tools on a more manageable scale. Where PTF produces around $10$ transients per night, LSST is expected to yield more than $10^3$ per night and orders of magnitude more variable stars \citep{2004AAS...20510812B}. In PTF these events are outnumbered 2--3 orders of magnitude by uninteresting candidates, and with the need for near-real-time discovery, an automated machine-driven discovery pipeline is clearly needed.

Under the ML paradigm (for further information about machine learning, we refer the interested reader to \citealt{bishop2006} and \citealt{Hastie09}), computer algorithms are designed to learn some set of concepts or relationships from observed data.  Typically, these inferred relationships are  exploited to predict some quantity of interest (e.g.,  the class) for new data.  As more data are collected, ML methods can continue to refine their knowledge about the data set, thereby strengthening the  predictions for future data.  Moreover, unlike humans, machine-learning methods can instantaneously and automatically produce statements about data  and can easily scale with growing data collection rates.  Presently, machine learning is receiving much attention in time-domain astronomy, in such diverse areas as variable star classification (\citealt{2011ApJ...733...10R}), quasar selection using variability metrics (\citealt{2011ApJ...735...68K}), real-time GRB follow-up (\citealt{2012ApJ...746..170M}), photometric supernova typing (\citealt{2012MNRAS.419.1121R}), and exoplanet signal processing (\citealt{2012MNRAS.419.2683G,2012ApJ...747...12W}).  For a review  of machine learning in astronomy, see  \citet{2010IJMPD..19.1049B}.

  \citet{bloom10} first used ML to perform discovery and classification of variability for the Palomar Transient Factory (PTF).  For each candidate, a set of heuristics meant to capture an object's validity were measured and a machine-learned classifier was employed to separate the true variable and transient objects from a haystack of false detections. While the results were promising and enabled the continuous and successful operation of PTF, they were based on a quick implementation on the very restricted, manually labelled training set that was available at the time. In this paper, we introduce a second-generation approach, aimed at developing a method that works well on current survey data and has properties that can scale to even larger surveys such as LSST.  Our sole purpose here is to robustly determine whether a source is \emph{real}, i.e., a bona fide astrophysical source, or what we will call \emph{bogus} (following \citealt{bloom10}), an artefact of no astronomical interest. In this paper we describe how to optimize the process, discuss the fundamental astrophysical, computational, and statistical questions in this endeavour, and demonstrate high levels of performance on real (PTF) data.

  The actual experimental procedure involves iteratively revisiting various steps in the process. For example, one cannot choose a classification algorithm without feeding it features, yet it is quite difficult to choose the best features without a classifier. However, for clarity, we linearise the analysis and description. 
  In the following section (\S\ref{s:classification}) we discuss building of the training set, extracting features from the data and how these are used to build the real-bogus (RB2) classifier. In order to optimize the model, we perform evaluation experiments in \S\ref{s:optim}, including the selection of an optimal subset of features and the tuning of model parameters. We present the application of this framework to PTF data (\S\ref{s:application}) and before we conclude in \S\ref{s:conclusion}, we investigate in \S\ref{s:contamination} the effect of contamination of the training and validation data with false labels, an important insight for future synoptic surveys.

\section{Machine-Learned Discovery through Supervised Classification}
\label{s:classification}

Supervised classification describes a set of ML methods that use a \emph{training set} of observed sources with known class membership to learn a function that describes the relationship between the observed data and the class to which a source belongs.  Once a suitable classification function (a.k.a. \emph{classifier}) is estimated, it can be employed to predict the class of any future object from its observed data.  Implemented as part of a framework, this effectively provides us with an automated classification engine, or in the case of the real-bogus classification problem, a discovery engine by which truly varying astrophysical objects can be separated from bogus detections.  Moreover, when properly trained and validated, these methods provide a statistical guarantee on the performance of the discovery engine for new data, meaning that we can be assured that the false positive and missed detection rates of the classifier lie in some narrow range with very high probability.

In this section, we discuss the major components of our real-bogus ML classifier.  Supervised classification is typically very sensitive to the training set that is used to train the model.  In \S\ref{ss:trainset} we describe the means by which we accrued a robust real-bogus training set of PTF detections in order to minimize sample-selection bias.  Second, the representation of the observed data is highly influential in the performance of our algorithms.  In \S\ref{ss:features} we describe our compression of  pixelated reference and subtraction  images into a set of real-valued \emph{features} which are devised to contain the relevant information content about a source's ``realness" while ignoring uninformative content.  And in \S\ref{ss:method} we describe the ML \emph{classifier}, which is a non-parametric statistical model, learned from data, that maps from the vector of observed features to the set of real/bogus classes.

  \subsection{Training Set of Transient Candidates}
  \label{ss:trainset}

In order to build a training set of candidates with known class labels at the beginning of a new survey one typically would have to rely on data from previous surveys, from simulations, or from limited (and often not representative) commissioning data. As we show below, the quality and sheer size of a training set have a tremendous impact on the robustness of the classification. In fact, most of the improvement we report here, compared to the success rates of \citet{bloom10} can be attributed to supplying a substantially larger training set (by 2 orders of magnitude) which is more representative of the population of PTF transient candidates. We  have the advantage of retrospect, with many months of data taking behind us, and a daease that includes more than a hundred million candidate sources. 

However, the challenge lies in the fact that most of these sources are spurious, and only a small minority have been scanned by humans in order to provide us with a ground truth. Having multiple domain experts scan the many millions of candidates is obviously impossible.   Fortunately, we can obtain an adequate real-bogus labelled sample via the following procedure:
\begin{itemize}
\item First, a sample of \emph{bogus} sources can be easily (though not perfectly) achieved by randomly selecting sources from our database and removing known variable sources. Since the majority of objects in the database are \emph{bogus}, the sample should have little contamination from \emph{real} sources (we explore this in detail in \S \ref{ss:rbratio}). This process generates 63,667 bogus sources for our PTF training sample cleaned from 70,000 randomly selected candidates.
\item Second, a sample of \emph{real} candidates can be obtained from a list of identified sources in our database. We limit the sample to objects discovered in 2010 (thus eliminating the first few months of commissioning, and data obtained before an electronics upgrade). We total $14,781$ individual detections of which $10,548$ are of $569$ individual supernovae, $1235$ are of Active Galactic Nuclei (AGN), and $2731$ are of variable stars. All of these sources were either identified by the collaboration spectroscopically (most) or had a classification in the public domain (such as Vizier; \citealt{vizier2000}). 
\end{itemize}
Though this selection procedure introduces some incorrect labels, we demonstrate in \S \ref{s:contamination} that label noise in the training set does not significantly hurt performance of the classifier, at least for the typical proportions of incorrectly labelled data expected in our training set.

This labelled sample is biased by the fact that we rely on the previous generation of PTF detections to identify sources. Since the primary focus of PTF is  supernova discovery, our sample is heavily skewed toward supernovae (two-thirds of our confirmed reals are supernovae), and is biased toward discovering objects that resemble those that have been found and followed up by the PTF collaboration. However, by including in our labelled set all the candidates associated spatially with a given real source, and not just the single detection that triggered the `discovery', we can expand our training sample to fainter, lower signal-to-noise (S/N) detections, and thereby extend the capabilities of our classifier beyond the reaches of the previous algorithm. For example, a confirmed SN may not have been identified as a promising candidate at early times, when it was relatively faint, despite being detected and tabulated.  Yet, we now can train on that detection, which permits us to increase our discovery power for fainter objects. Section \S\ref{ss:deeper} goes into more details on this.

Our selection procedure yields 78,448 labelled PTF detections: 14,781 real and 63,667 bogus detections.  In order to train and validate our classifier, we divide this labelled data set into disjoint sets of training examples, $\mathcal{T}$, and validation examples, $\mathcal{V}$. Under this paradigm, our algorithm learns an appropriate classifier using the data examples in $\mathcal{T}$ and then evaluates its performance against the data examples in $\mathcal{V}$. Training the classifier on a set $\mathcal{T}$ and verifying against a disjoint set $\mathcal{V}$ is a necessary practice when building a machine learner because it allows us to justify that our classifier has ``learned'' the intrinsic properties of the problem rather than being overly specialized (and over-fitted) to the training set $\mathcal{T}$. 

For validation on PTF data in Section \ref{s:application} we split the labelled set into a set of 50,000 training and 28,448 validation examples.  This split is performed randomly, with the caveat that `paired' observations are always placed into the same set.  These paired observations arise because every field in PTF is typically observed twice during a night with a time separation of about an hour. This observing strategy is employed by PTF to find and reject asteroids, the vast majority of which will move a detectable amount with respect to the observer frame between observations. For a total of 5,236 of detections in our labelled set, we also have a paired candidate that was detected at the same position (within 0.001 degrees) within 2.5 hours.  By always placing all paired objects in the same set (either in $\mathcal{T}$ or $\mathcal{V}$), we ensure that the performance of our classifier is not artificially over-stated due to training on sources whose counterparts, observed on the same night within a few hours of one another, are also used for validation.

  \subsection{Feature Representation of Candidates}
  \label{ss:features}

\begin{figure}
\centering
  \includegraphics[width=0.8\columnwidth,angle=0]{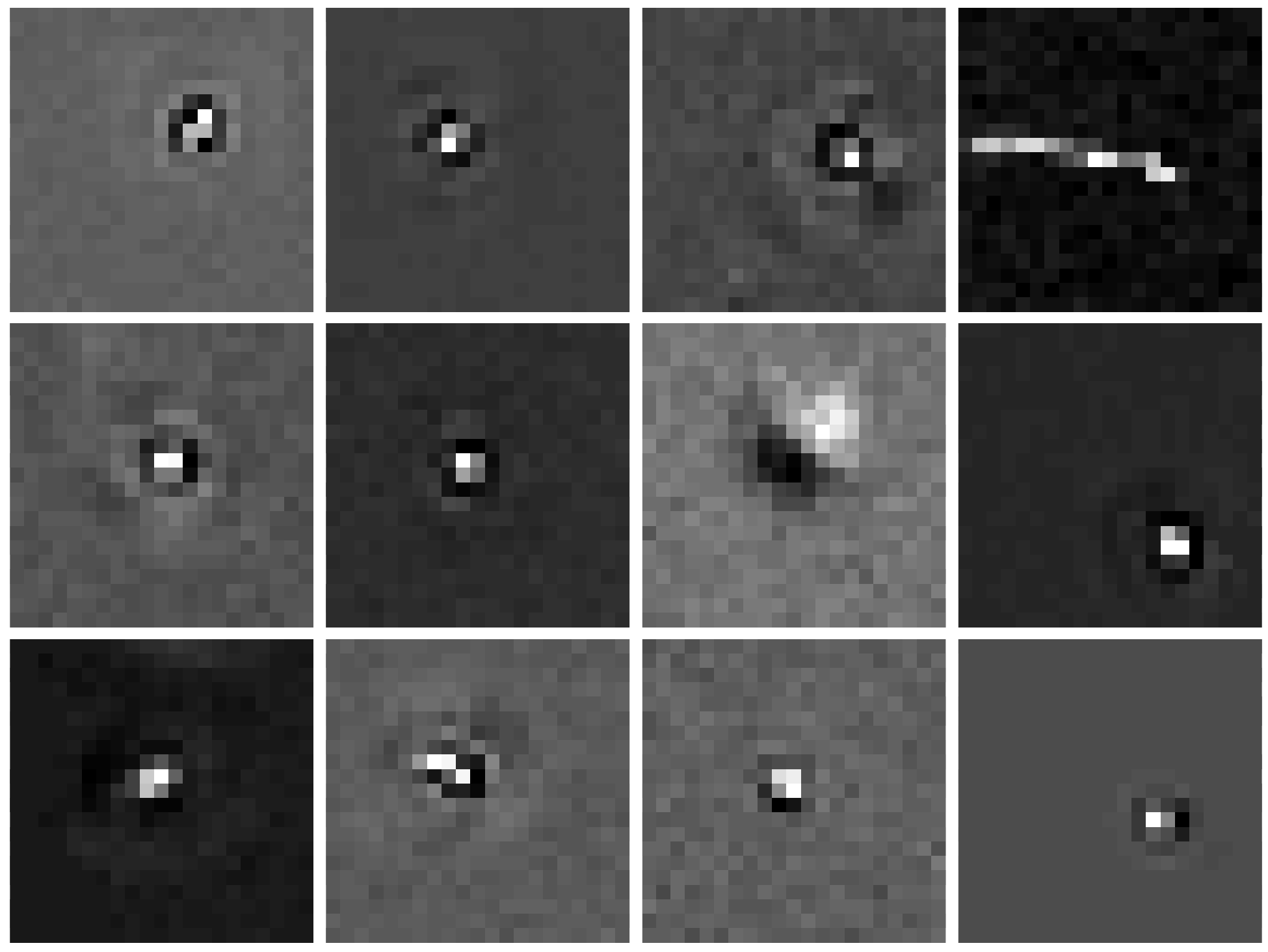}
	\begin{minipage}{0.8\columnwidth} 
	\vspace{-0.5in}
	\hspace{0.1in}
	\end{minipage}
  \includegraphics[width=0.8\columnwidth,angle=0]{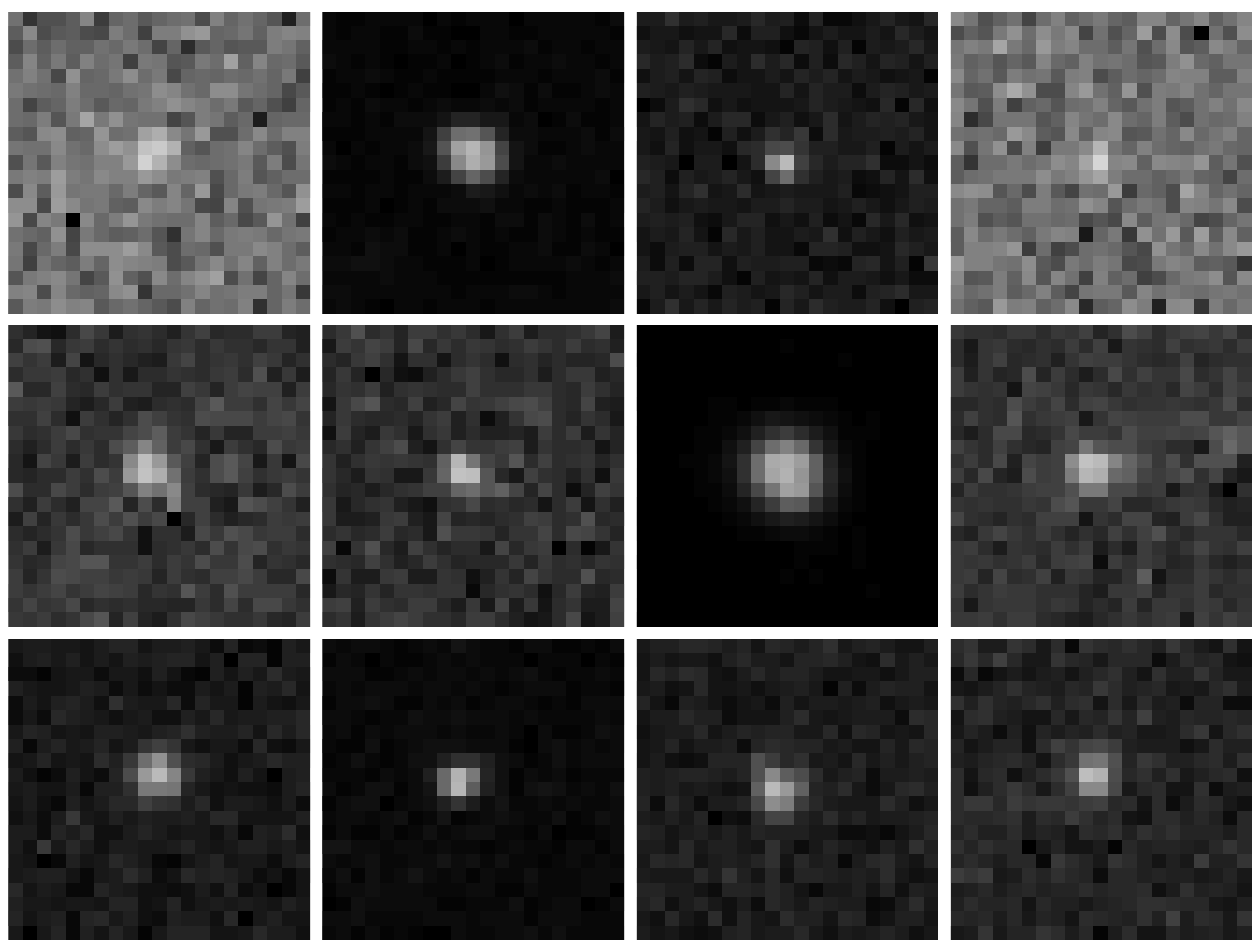}
  \caption{Examples of \emph{bogus} (top) and \emph{real} (bottom)
    thumbnails. Note that the shapes of the \emph{bogus} sources can be quite varied, which poses
    a challenge in developing features that can accurately represent
    all of them.  In contrast, the set of \emph{real} detections is more uniform in terms of the
    shapes and sizes of the subtraction residual.  Hence, we focus on finding a compact set of features that
    accurately captures the relevant  characteristics of \emph{real} detections
    as discussed in \S\ref{ss:features}.}
  \label{f:ex}
\end{figure}

    The role of a feature set is to provide a succinct representation of each candidate that captures the salient class information encoded in the observed data while discarding useless information.  Determining a useful set of features is critical because the classifier relies on this information to learn the proper feature--class relationship in order to perform effective classification.  The challenge, then, is to determine which are the features that (a) can be easily and quickly derived from the available data for each candidate and (b) can be used to effectively separate the real from the bogus candidates.    For every real or bogus candidate, we have at our disposal the subtraction image of the candidate (which is reduced to a 21-by-21 pixel---about 10 times the median seeing full width at half maximum---\emph{postage stamp} image centered around the candidate), and metadata about the reference and subtraction images.  Figure \ref{f:ex} shows  subtraction thumbnail images for several arbitrarily chosen bogus and real candidates.

     In this work, we supplement the set of features developed by \citet{bloom10} with image-processing features extracted from the subtraction images and summary statistics from the PTF reduction pipeline.  These new features---which are detailed below---are designed to mimic the way humans can learn to distinguish real and bogus candidates  by visual inspection of the subtraction images. For convenience, we describe the features from \citet{bloom10}, hereafter the RB1 features, in Table~\ref{t:RBfeats}, along with the features added in this work. In \S\ref{ss:featselect}, we critically examine the relative importance of all the features and select an optimal subset for real--bogus classification.

Prior to computing features on each subtraction image postage stamp, we normalize the stamps so that their pixel values lie between $-1$ and $1$.  As the pixel values for real candidates can take on a wide range of values depending on the astrophysical source and observing conditions, this normalization ensures that our features are not overly sensitive to the peak brightness of the residual nor the residual level of background flux, and instead capture the sizes and shapes of the subtraction residual.  Starting with the raw subtraction thumbnail, $I$, normalization is achieved by first subtracting the median pixel value from the subtraction thumbnail and then dividing by the maximum absolute value across all median-subtracted pixels via
    \begin{equation}
    \label{e:trans}
      I_\mathrm{N}(x,y) = \left\{ \frac{I(x,y)-\mathrm{med}[I(x,y)]}{\mathrm{max}\{\mathrm{abs}[I(x,y)]\}} \right\} .
    \end{equation}
    Analysis of the features derived from these normalized real and bogus subtraction images showed that the transformation in (\ref{e:trans}) is superior to other alternatives, such as the Frobenius norm ($\sqrt{\mathrm{trace}(I^T I)}$) and truncation schemes where extreme pixel values are removed.

    Using Figure \ref{f:ex} as a guide, our first intuition about real candidates is that their subtractions are typically azimuthally symmetric in nature, and well-represented by a 2-dimensional Gaussian function, whereas bogus candidates are not well behaved.  To this end, we define a spherical 2D Gaussian, $G(x,y)$, over pixels $x,y$ as
        \begin{equation}
      G(x,y) = A \cdot \exp \left\{-\frac{1}{2} \left[ \frac{(c_x - x)^2}{\sigma} + \frac{(c_y - y)^2}{\sigma} \right] \right\} ,
    \end{equation}
which we fit to the normalized PTF subtraction image, $I_N$, of each candidate by minimizing the sum-of-squared difference between the model Gaussian image and the candidate postage stamp with respect to the central position $(c_x, c_y)$, amplitude $A$\footnote{As  subtraction images of real candidates can be negative when the brightness of the source is decreasing, we allow the Gaussian amplitude $A$ to take on negative, as well as positive, values.} and scale $\sigma$ of the Gaussian model.   This fit is obtained by employing an \emph{L-BFGS-B} optimization algorithm \citep{byrd1995}. The best fit scale and amplitude determine the \texttt{scale} and \texttt{amp} features, respectively, while the \texttt{gauss} feature is defined as the sum-of-squared difference between the optimal model and image, and \texttt{corr}  is the Pearson correlation coefficient between the best-fit model and the subtraction image.  

Next, we add the feature \texttt{sym} to measure the symmetry of the subtraction image.  The \texttt{sym} feature should be small for real candidates, whose subtraction image tends to have a spherically symmetric residual. \texttt{sym} is computed by first dividing the subtraction thumbnail into four equal-sized quadrants, then summing the flux over the pixels in each quadrant (in units of standard deviations above the background) and lastly averaging the sum-of-squares of the differences between each quadrant to the others. Thus, \texttt{sym} will be large for difference images that are not symmetric and will be nearly zero for highly symmetric difference images.
    
    Next, we introduce features that aim to capture the smoothness characteristics of the subtraction image thumbnails. A typical real candidate will have a smoothly varying subtraction image with a single prominent peak while bogus candidates generally have multiple peaks and more complex structure.  To capture this behaviour, we introduce the feature \texttt{l1}, defined as
       \begin{equation}
    \label{e:l1}
      \ell_1(I_N) = \frac{\sum |I_N(x,y)|}{\sqrt{\sum I_N(x,y)}},
    \end{equation}
    which is the $\ell_1$ norm of the normalized image.  This feature measures the relative sparsity of the image, that is, the number of pixels that have quite large count values relative to the others; a real candidate should have relatively few such pixels. Additionally, we compute the features \texttt{smooth1} and \texttt{smooth2}, which capture the maximum pixel value after passing each subtraction image through a  $3 \times 3$ and $5 \times 5$ moving average and difference filter, respectively.  These filters are designed with $1$'s everywhere except the center pixel, which is set to $-1$.  Thus, when convolved with an image, the filter averages each small square of the image while subtracting off the center pixel. These features are constructed to capture structure on the scale of the typical size of a real candidate subtraction, so we expect a larger number from the convolution of these kernels on real candidates.

    As part of the exploratory analysis of the imaging data, we employed a principal component analysis (PCA) of the postage stamp subtraction images. PCA is a useful image analysis tool as the eigenvectors of the covariance matrix have the same dimension as the images themselves and can be plotted to show typical example images. We use the projection of each postage stamp subtraction image on the first two principal components of the full set of images to provide us with two PCA features, \texttt{pca1} and \texttt{pca2}.

   Finally, we add a few contextual features from the PTF image subtraction process.  The identification number of the CCD chip, \texttt{ccid}, is included to aid in discovery if there are cosmetic differences between the arrays that can cause some to have more artefacts than others. Along the same line, \texttt{extracted} and \texttt{obsaved}, the total number of candidates detected and saved by {\tt SEXTRACTOR} \citep{bertin96}, respectively, quantify the quality of the exposure:  the higher the number of sources found or extracted, the higher the likelihood they are bogus. Next, \texttt{seeingnew}, the FWHM of seeing for the new exposure, can help identify exposures with poor image quality. Lastly, we add \texttt{pos}, an indicator of whether the image subtraction has mostly positive or negative residual pixels, to separate candidates which have brightened compared to the reference, from those that have dimmed, which may have distinct observational characteristics.  
       
    \begin{figure*}
      \centering
    \includegraphics[width=\textwidth,angle=0]{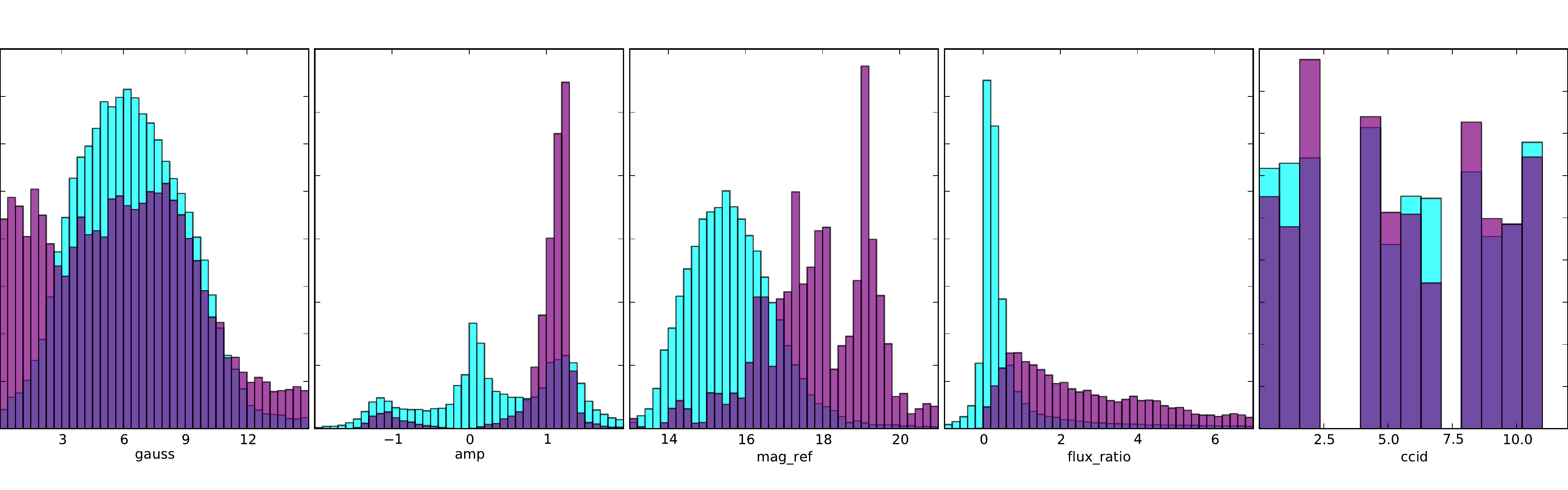}
    \caption{Histograms of training set objects for a selection of five of the most important classification features, divided into their real (purple) and bogus (cyan) populations. From left to right, \texttt{gauss}, which is the goodness-of-fit of the Gaussian fit, \texttt{amp}, the amplitude of that  fit, \texttt{mag\_ref}, the magnitude of the source in the reference image, \texttt{flux\_ratio}, the ratio of the fluxes in the new and reference images and lastly, \texttt{ccid}, the ID of the camera CCD in which the source was detected. The fact that this last feature is useful is somewhat surprising, but we can clearly see that on certain CCDs, the probability that a candidate observed on that chip has a different conditional likelihood of being a real source.}
      \label{f:comparemanyfeatures}
    \end{figure*}

    This brings us to a total of 42 features, which are summarized in Table \ref{t:RBfeats}. To visualize the ability of these features to separate real from bogus candidates, we plot histograms of a few of the most discriminating features in Figure \ref{f:comparemanyfeatures}. This shows the separation potential in these dimensions individually, but we will now turn our attention to building an effective machine-learned classifier, where the challenge is to determine a decision boundary in the 42-dimensional space spanned by all features. We can calculate features for around 2 candidates per second in serial and trivially parallelize as candidates are independent.

\begin{table*} 
\begin{minipage}{185mm}
\begin{tabular}{l|cll}
Set & Selected & Feature & Description\\
\hline 
RB1 &  &\texttt{mag}     & USNO-B1.0 derived magnitude of the candidate on the difference image \\
&  &\texttt{mag\_err} & estimated uncertainty on \texttt{mag} \\
&  &\texttt{a\_image} & semi-major axis of the candidate\\
& \checkmark &\texttt{b\_image} & semi-minor axis of the candidate \\
&  & \texttt{fwhm}    & full-width at half maximum (FWHM) of the candidate \\
& \checkmark & \texttt{flag}    & numerical representation of the SExtractor extraction flags \\
& \checkmark & \texttt{mag\_ref} & magnitude of the nearest object in the reference image if less than \\ & & & 5 arcsec from the candidate \\
& \checkmark & \texttt{mag\_ref\_err} & estimated uncertainty on \texttt{mag\_ref} \\
& \checkmark & \texttt{a\_ref}  & semi-major axis of the reference source \\
& \checkmark & \texttt{b\_ref}  & semi-minor axis of the reference source \\
&  & \texttt{n2sig3} & number of at least negative 2 $\sigma$ pixels in a 5$\times$5 box centered on the candidate \\
&  & \texttt{n3sig3} & number of at least negative 3 $\sigma$ pixels in a 5$\times$5 box centered on the candidate \\
&  & \texttt{n2sig5} & number of at least negative 2 $\sigma$ pixels in a 7$\times$7 box centered on the candidate \\
& \checkmark & \texttt{n3sig5} & number of at least negative 3 $\sigma$ pixels in a 7$\times$7 box centered on the candidate \\
& \checkmark & \texttt{flux\_ratio} & ratio of the aperture flux of the candidate relative to the aperture flux \\ & & & of the reference source \\
&  & \texttt{ellipticity} & ellipticity of the candidate using \texttt{a\_image} and \texttt{b\_image} \\
& \checkmark & \texttt{ellipticity\_ref} & ellipticity of the reference source using \texttt{a\_ref} and \texttt{b\_ref} \\
& \checkmark & \texttt{nn\_dist\_renorm} & distance in arcseconds from the candidate to reference source \\
&  & \texttt{magdiff} & when a reference source is found nearby, the difference between the candidate\\ & & & magnitude and the reference source. \\ & & & Else, the difference between the candidate  magnitude \\ &&& and the limiting magnitude of the image\\
& \checkmark & \texttt{maglim} & True if there is no nearby reference source, False otherwise.\\
&  & \texttt{sigflux} & significance of the detection, the PSF flux divided by the  \\ & & & estimated uncertainty in the PSF flux \\
&  & \texttt{seeing\_ratio} & ratio of the FWHM of the seeing on the new image to the FWHM  \\& & &  of the seeing on the reference image \\
& \checkmark & \texttt{mag\_from\_limit} & limiting magnitude minus the candidate magnitude \\
&  & \texttt{normalized\_fwhm} & ratio of the FWHM of the candidate to the seeing in the new image \\
& \checkmark & \texttt{normalized\_fwhm\_ref} & ratio of the FWHM of the reference source to the seeing in the \\& & & reference image \\
& \checkmark & \texttt{good\_cand\_density} & ratio of the number of candidates in that subtraction to the total \\& & & usable area on that array \\
& \checkmark & \texttt{min\_distance\_to\_edge\_in\_new} & distance in pixels to the nearest edge of the array on the new image \\
\hline 
New & \checkmark &    \texttt{\textbf{ccdid}} & numerical ID of the specific camera detector ($1-12$)\\
&   &     \texttt{sym} & Measure of symmetry, based on dividing the object into quadrants \\
& \checkmark  &     \texttt{seeingnew} & FWHM of the seeing on the new image\\
& \checkmark  &     \texttt{extracted} & number of candidates on that exposure found by Sextractor\\
& \checkmark  &     \texttt{obsaved} &  number of candidates on that exposure saved to the database (a subset of \texttt{extracted})\\
&   &     \texttt{pos} & True for a positive (i.e., brighter) residual, False for a negative (fading) one\\
& \checkmark  &     \texttt{gauss} & gaussian best fit sqaured difference value \\
&   &     \texttt{corr} & gaussian best fit correlation value \\
&   &     \texttt{scale} & gaussian scale value\\
& \checkmark  &     \texttt{amp} & gaussian amplitude value\\
& \checkmark  &     \texttt{l1} & sum of absolute pixel values\\
&   &     \texttt{smooth1} & filter $1$ output\\
&   &     \texttt{smooth2} & filter $2$ output\\
&   &     \texttt{pca1} & 1st principal component\\
&   &     \texttt{pca2} & 2nd principal component\\
\hline
Test & &    \texttt{empty} & zero for all candidates (i.e., no information)\\
  &  &   \texttt{random} & a random number generated for every candidate (i.e., pure noise)\\
  \hline
 \end{tabular}
\end{minipage}
\caption{List of all of the features used in our analysis. The first set of features, labeled `RB1', were first introduced by \citet{bloom10} and we repeat here their Table 1. The second, labeled `New' is introduced here. The last set of features, called `Test' serves as a benchmark for feature selection in \S\ref{ss:featselect}, where we expect good features to perform better than these.  The check-marked as `selected' represent the optimal subset found by our incremental feature selection algorithm in \S\ref{ss:featselect}.\label{t:RBfeats}} 
\end{table*}

  \subsection{Random Forest Supervised Classification}
  \label{ss:method}
  
    There are a number of methods that one can use for supervised classification. For example, support vector machines, logistic regression, boosting, decision trees, and random forests have all experienced wide use in statistics and machine learning (for details and examples see \citealt{Hastie09}).  Previously, \citet{2007ApJ...665.1246B} compared many ML classifiers for supernova search. In the present work, we employ random forest classification, which has shown high levels of performance in the astronomy literature (e.g., \citealt{2010ApJ...712..511C,2011ApJ...733...10R, 2011MNRAS.414.2602D}). A description of the algorithm  can be found in \citet{Brei01}. Briefly, the method aggregates a collection of hundreds to thousands of classification trees, and, for a given new candidate, outputs the fraction of classifiers that vote \emph{real}. If this fraction is greater than some threshold $\tau$, then the random forest classifies the candidate as \emph{real}; otherwise it is deemed \emph{bogus}. 

  While an ideal classifier will have no missed detections (i.e., no \emph{real} identified as \emph{bogus}), with zero false positives (\emph{bogus} identified as \emph{real}), a realistic classifier will typically offer a trade-off between the two types of errors. A receiver operating characteristic (ROC) curve is a commonly used diagram which displays the missed detection rate (MDR) versus the false positive rate (FPR) of a classifier\footnote{Note that the standard form of the ROC is to plot the false positive rate versus the true positive rate (TPR = 1-MDR)}. With any classifier, we face a trade-off between MDR and FPR: the larger the threshold $\tau$ by which we deem a candidate to be real, the lower the MDR but higher the FPR and vice versa.  Varying $\tau$ maps out the ROC curve for a particular classifier, and we can compare the performance of different classifiers by comparing their cross-validated ROC curves: the lower the curve the better the classifier. 
  
  A commonly used figure of merit (FoM) for selecting a classifier is the so-called Area Under the Curve (AUC, \citealt{friedman2001elements}), by which the classifier with minimal AUC is deemed optimal.  This criterion is agnostic to the actual FPR or MDR requirements for the problem at hand, and thus is not appropriate for our purposes.  Indeed, the ROC curves of different classifiers often cross, so that performance in one regime does not necessarily carry over to other regimes.  In the real--bogus classification problem, we instead define our FoM as the MDR at 1\% FPR, which we aim to minimize.  The choice of this particular value for the false positive rate stems from a practical reason: we do not want to be swamped by bogus candidates misclassified as real. 

\begin{figure}
      \centering
    \includegraphics[width=\columnwidth,angle=0]{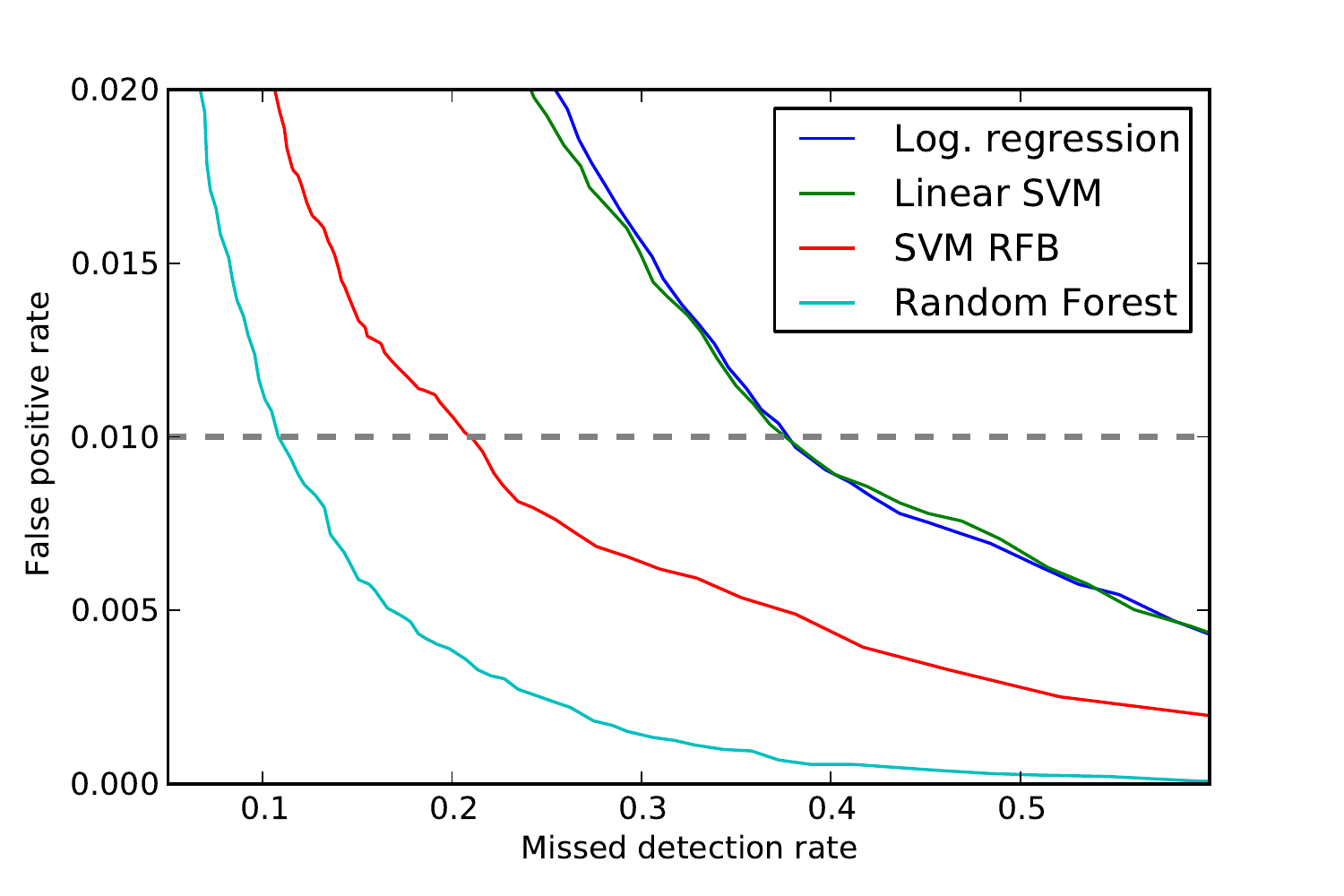}
     \caption{Comparison of a few well known classification algorithms applied to the full dataset. ROC curves enable a trade-off between false positives and missed detections, but the best classifier pushes closer towards the origin. Linear models (Logistic Regression or Linear SVMs) perform poorly as expected, while non-linear models (SVMs with radial basis function kernels or random forests) are much more suited for this problem. Random forests perform well with minimal tuning and efficient training, so we will use those in the remainder of this paper.}
      \label{f:compare_rocs}
    \end{figure}

  Figure \ref{f:compare_rocs} shows example ROC curves comparing the performance on pre-split training and testing sets including all features. With minimal tuning, random forests perform better, for any position on the ROC curve, than SVM with a radial basis kernel, a common alternative for non-linear classification problems. A line is plotted to show the 1\% FPR to which our figure of merit is fixed.

\section{Optimizing the Discovery Engine}
\label{s:optim}

With any machine learning method, there are a plethora of modelling decisions to make when attempting to optimize predictive accuracy on future data.  Typically, a practitioner is faced with questions such as which learning algorithm to use, what subset of features to employ, and what values of certain model-specific tuning parameters to choose. Without rigorous optimization of the model, performance of the machine learner can be hurt significantly.  In the context of real--bogus classification, this could mean failure to discover objects of tremendous scientific impact. In this section, we describe several choices that must be made in the real--bogus discovery engine and outline how we choose the optimal classification model by minimizing the cross-validated FoM over several different tuning parameters.

  \subsection{Feature selection}
  \label{ss:featselect}
    
    The features described in \S\ref{ss:features} all provide some level of discrimination between \emph{real} and \emph{bogus} sources, as seen in figure \ref{f:comparemanyfeatures}. However, having too many features will often hinder the statistical performance of the classifier as well as increase the computational complexity of the resulting learning algorithm.  Thus, the next modelling decision with which we are faced is {\it whether to select a subset of features to use in the classifier and if so, choosing the optimal subset to use}.  Feature selection can serve multiple purposes, including to increase the interpretability of our classifications, improve the computational performance, and strengthen the quality of the classification, as measured by the FoM. With a fixed number of training data samples,  statistical prediction is often made more difficult with an increasing number of features, owing to the danger of over-fitting and the difficulty of estimating the proper decision boundaries in high-dimensional feature spaces. As we show below, removing some features can improve the real--bogus figure of merit.

    Feature selection is in general a very challenging problem. Our goal is to find a subset of the $n$ features that best explains the data, so we would ideally want to experiment on any subset of features of any size $s$. For a realistic number of features this is computationally intractable, as we would have to train $\sum_{s=1}^{n-1} {n \choose s}$ classifiers. 

    Instead we have to rely on approximation methods for finding the best subset of features. Well-known variations of such methods are the forwards and backwards feature selection algorithms \citep{guyon2003}. The forward selection algorithm starts with an empty set of features and iteratively adds features that improve the classifier the most, while the backwards selection method starts with the full set of features and iteratively removes features that improves the classifier the most or hurts the classifier the least. As the random forest classifier with our choice of figure of merit does not work well on very small sets of features, we choose the backwards selection method for this experiment.

    As hinted above, the backwards selection method starts out with the full set of features and aims to remove features that provide no useful information, or actually hurt the classification performance. We do this by calculating a 5-fold cross-validated figure of merit for all models with a single feature removed. The model with the lowest figure of merit is then selected, effectively removing the associated feature, and we iterate the process until some criterion is satisfied. In this case we stop when our figure of merit is no longer defined when the ROC curve no longer goes below a false-positive rate of $1\%$ with only a few features left in the model. This procedure requires at most $ \sum_{i=1}^{N-1} N-i $ steps. Choosing 20 features from a set of 40 we go through $610$ steps, while the exhaustive search would have required ${40 \choose 20} \sim 10^{11}$.

    \begin{figure}\centering
      \includegraphics[width=\columnwidth]{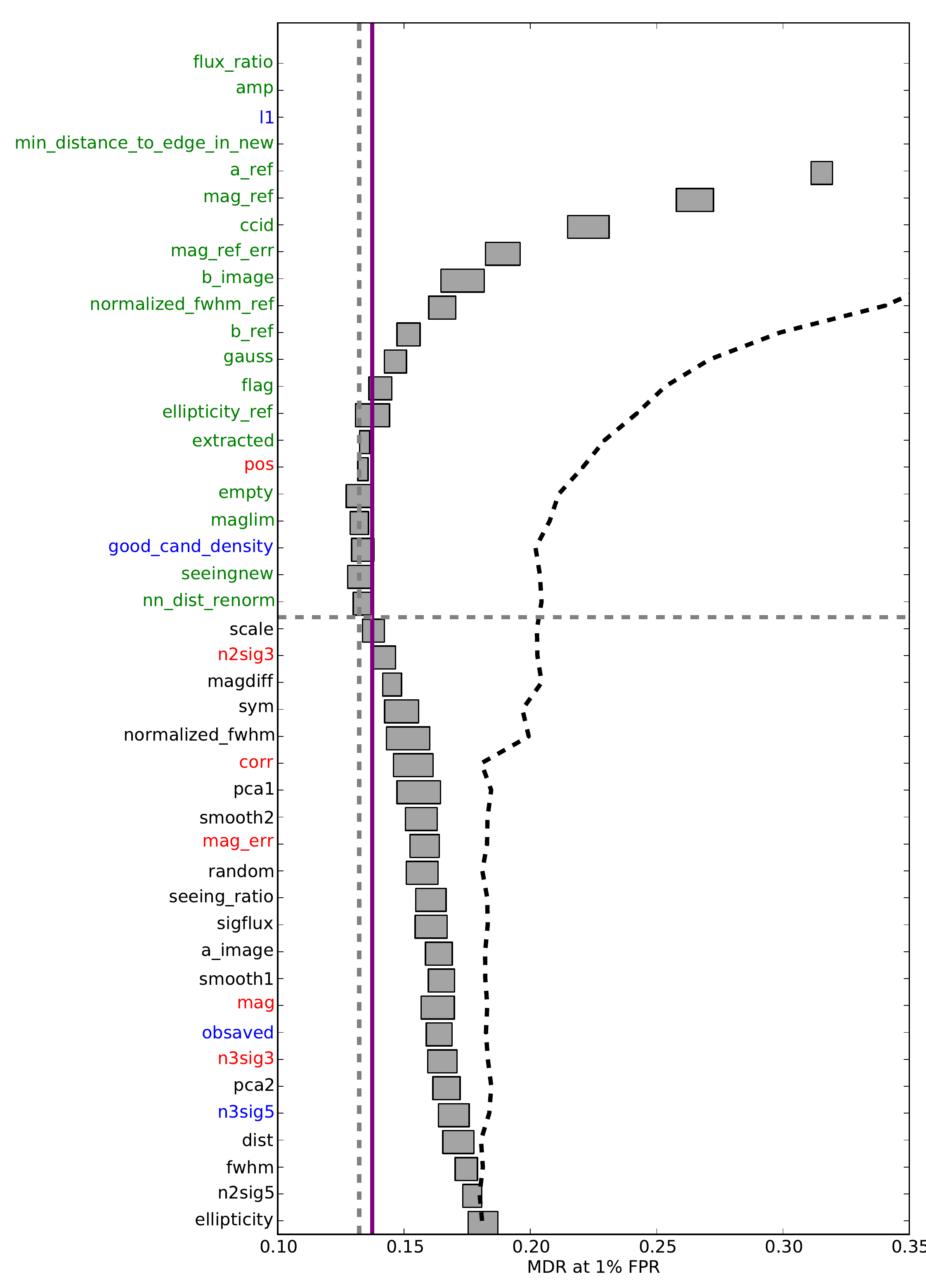}
      \caption{Backwards feature selection. We start with the full model at the bottom and iteratively eliminate the feature whose absence yields the best 5-fold cross-validated figure of merit. Boxes show the cross-validated mean $\pm \sigma$ for the model without the corresponding feature. The vertical dashed line shows the optimal FoM, the vertical purple line shows the threshold FoM (minimum + $1 \sigma$) and features above the dashed horizontal line are below the threshold and selected for the optimal model. This procedure is rerun 5 times and feature names are coloured by the number of times they are selected: 0 (black), 1 (red), 2 (blue) and 3+ (green). The dashed black line shows the result of iteratively removing features based on random forest feature importance, but note that this line does not follow the feature names on the y-axis.}
      \label{f:feature_selection}
    \end{figure}

    In Figure \ref{f:feature_selection} we see the results of the backwards selection process. In contrast to \citet{biau2010}, where random forests are shown to be relatively immune to noisy features, we see in figure \ref{f:feature_selection} an improvement in our particular cross-validated figure of merit---by more than 4\% in terms of MDR---by removing features. In some other cases of real-world datasets, random forests have been shown to over-fit to the training set because of the un-pruned trees \citep{segal2004}, which might be what is happening here. The random forest tuning parameters {\tt ntree} and {\tt nodesize} can minimize overfitting in the algorithm, but varying these showed little effect on the performance of the model on all features as well as on the optimal set of features as seen in section \S\ref{ss:tuning}
    
    Another point to draw from figure \ref{f:feature_selection} is that this feature selection method is relatively robust and does not change the ordering of features considerably between runs. Most of the selected features are always selected, while a few are only selected a few times. These occasionally selected features are usually exchanged with other features with which they are highly correlated, supporting the intuition that this feature selection method will try to find the optimally uncorrelated set of features. For the final model, we select all features that have been selected at least twice. This leaves us with 23 features from both \citet{bloom10} and section \S\ref{ss:features} in this paper, check-marked in Table~\ref{t:RBfeats}.

    To get a better handle on the feature selection process, we have in the above experiments introduced two benchmark features: \texttt{empty} and \texttt{random}. The former inserts a column of constant value (all $0$) into the dataset, in essence a feature without any useful information. The latter inserts a pure noise column of uniform random numbers at an attempt to hurt the classifier. The benchmark is to see at which point these features are removed from the feature selection process. As can be seen from figure \ref{f:feature_selection}, the random feature is removed relatively early, while the empty feature actually survives until the turnaround in the figure of merit which is somewhat unexpected. This behaviour is probably due to the fact that the features removed in the first half of the selection process actually hurt the classifier (in terms of FoM). This includes the \texttt{random} feature and features that are highly correlated with other selected features. The \texttt{empty} feature is (by rule) never actually used as a splitting feature in any of the decision trees of the random forest algorithm, remaining neutral while the other hurtful features are removed. The literature on feature selection in the context of random forest is very sparse, and getting a better understanding of this behaviour is of great interest for future iterations of our classifier and related projects.
    
    An alternative way of selecting features, in the context of random forests, is to utilize the built-in feature importance metric of the RF classifier \citep{Brei01}. A central property of random forests is that features are left out of the model at random. This enables the algorithm to determine how the training error is affected when certain features have been left out, and thereby determining the relative importance of all features. To utilize this for feature selection, we employ a type of backwards selection based on feature importance where the feature with the least importance is removed iteratively until no features are left. The advantage of this method compared to the heuristics introduced above is that it is faster and built-in to the classification algorithm, but the disadvantage is that this method does not handle the correlation between features well \citep{strobl2008}. In our tests, this feature selection method does  not improve on the cross-validated classification performance of the random forest model with all features included, staying constant at $\sim 17\%$ until falling off when too many features are removed. The benchmark features \texttt{empty} and \texttt{random} are both removed early in this process.

  \subsection{Tuning the Random Forest}\label{ss:tuning}

Any non-parametric classification method typically brings with it a set of \emph{tuning parameters} which set the flexibility and adaptiveness of the model.  Random forest is no different, as it contains three important tuning parameters: (1) {\tt ntree}, the number of decision trees that compose the ensemble, (2) {\tt mtry}, the number of features that are randomly selected as splitting candidates in each non-terminal tree node, and (3) {\tt nodesize}, the size of a tree's terminal node, in terms of number of training objects populating the node, at which further feature splitting is disallowed.  Broadly, these tuning parameters act as smoothing parameters, affecting the complexity of the decision boundaries that are estimated by the random forest.  Optimal choice of these parameters depends on the complexity of the true real--bogus decision boundary in high-dimensional feature space.  Thus, our next modelling choice is to {\it determine which values of the tuning parameters produce the optimal real--bogus random forest classifier}.

\begin{figure}\centering
      \includegraphics[width=\columnwidth]{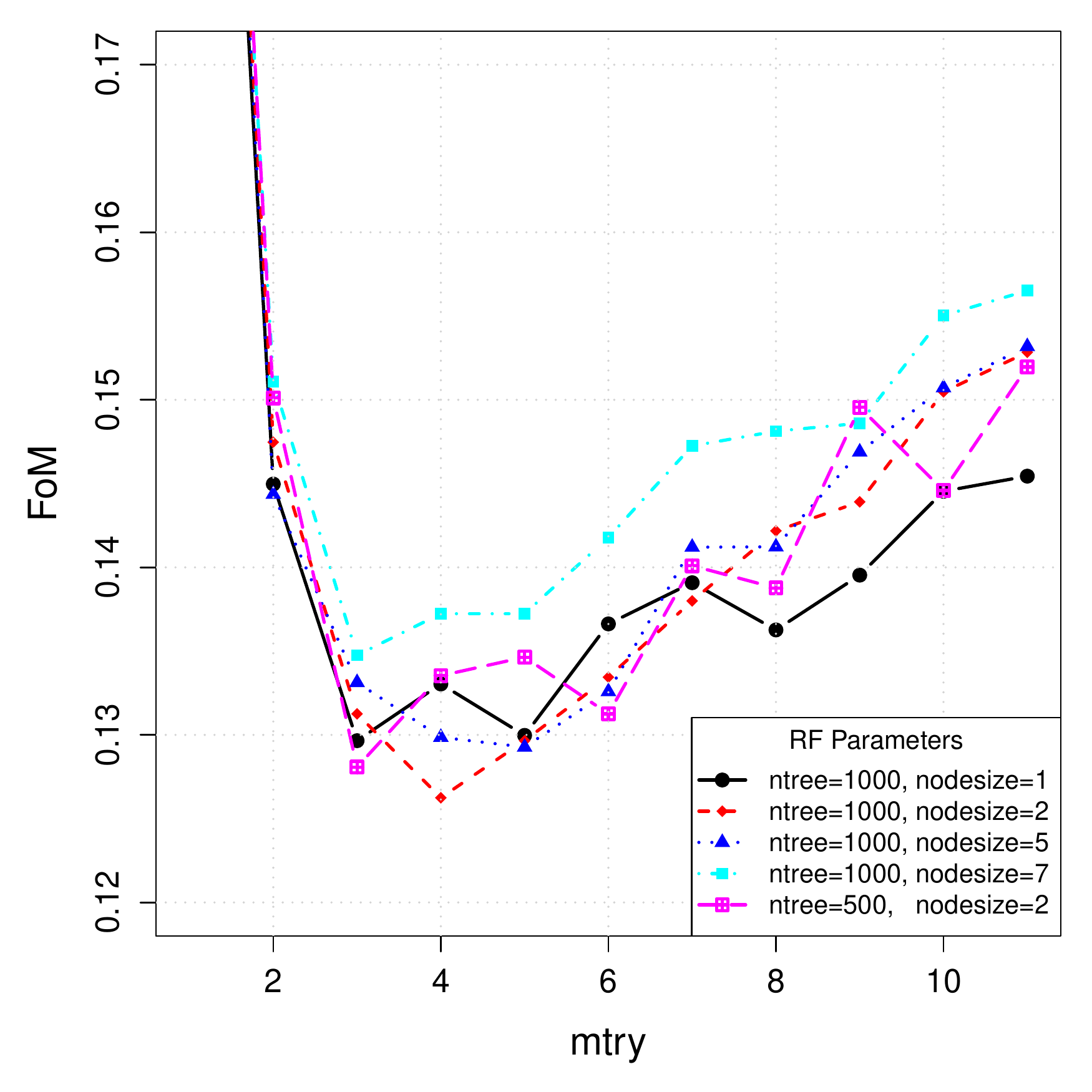}
      \caption{ Result of a grid search over the three random forest tuning parameters.  Each FoM is computed by averaging the individual FoMs for 10 random training--testing splits.  The tuning parameter {\tt mtry} is on the x-axis, while a separate line denotes the performance for different combinations of {\tt ntree} and {\tt nodesize}.  The optimal model is found to be {\tt ntree}=1000, {\tt mtry}=4, and {\tt nodesize}=2, though the performance is shown to be insensitive to small changes in the values of those parameters. \label{f:tuneparam} }
    \end{figure}

Using the optimal subset of features from \S\ref{ss:featselect}, we perform a grid search over the random forest tuning parameters {\tt ntree}, {\tt mtry}, and {\tt nodesize}.  Averaging over 10 random iterations of training--testing sets, we find that the average cross-validated FoM is minimized for the model ({\tt ntree}=1000, {\tt mtry}=4, {\tt nodesize}=2).  Moreover, in Figure \ref{f:tuneparam}, we find that the FoM is relatively insensitive to small changes in the values of the tuning parameters near the optimum.  For example, values of {\tt mtry} ranging from 3--6 each result in a cross-validated FoM within 1\% of the optimal solution.  The shape of the FoM curves with respect to {\tt mtry} is typical for non-parametric classifiers: for small values, the classifier is over-smoothed, resulting in high bias and low variance whereas for large values, the classifier is under-smoothed resulting in low bias and high variance; the optimal bias-variance trade-off is achieved somewhere in between.

      In a random PTF sample, the number of true \emph{bogus} candidates is much larger than the true \emph{real} candidates and we say that our data is \emph{imbalanced}. The issue that arises is that a classifier designed to minimize the probability of error can simply label everything as \emph{bogus}. As a consequence the probability of error will be of order $10^{-2}$, which is technically excellent but useless in practice. We circumvent this by building much less imbalanced training and validation sets, with a ratio of \emph{real} to \emph{bogus} of about one to four. Also, by optimizing the classifier for a specific point on the ROC curve (MDR at 1\% FPR), we specify the specific fraction of false positives and false negatives that is allowed. In \S \ref{ss:rbratio} we show how this choice of figure of merit affects the number of bogus candidates that would be presented after a certain RB2 threshold $\tau$.
      
  \subsection{Performance of optimized RB2 classifier}
  \label{ss:evaluation}
   
   In the previous sections we determined the optimal set of features and the optimal values of the random forest tuning parameters for the real--bogus problem. Although this method is ultimately an iterative process where one experiment will affect the other, we have constructed our final real--bogus model by tuning the random forest parameters after selecting the optimal subset of features. 

\begin{figure}\centering
      \includegraphics[width=0.9\columnwidth]{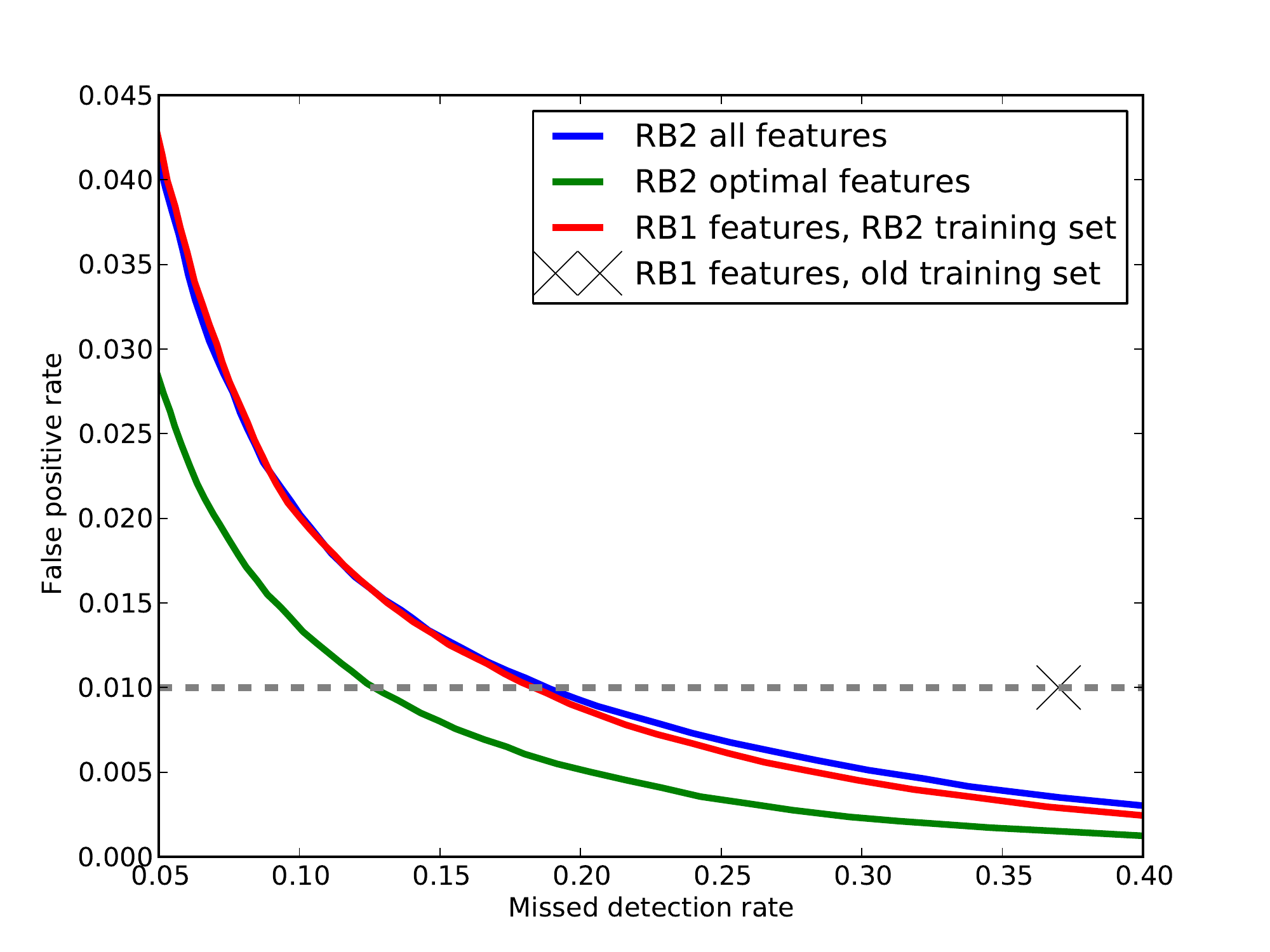}
      \caption{ROC curves showing the improvements in classification performance of RB2. The X marks the performance of RB1, and the red line shows the performance of the RB1 features with the new (much larger) training set. The blue line shows the little improved performance of the RB2 features on the new dataset, while selecting the optimal features yields the green line for a missed detection rate of $\sim 12\%$ at a $1 \%$ false-positive rate.}
      \label{f:performance}
    \end{figure}

    In order to isolate the amount of improvement in the classification performance that occurs due to the large increase in the amount of training data, versus that occurring due to the introduction and rigorous selection of new features, we compare the performance of three different classifiers: (1) a classifier built only with RB1 features but with the new training set, (2) the classifier constructed on all features, old and new, and (3) the classifier built on the optimal set of features yielded by our feature selection method in section \S\ref{ss:featselect}.  Results of this experiment  are plotted in Figure \ref{f:performance}, showing an overall improvement in the figure of merit from $\sim 35\%$ for the original RB1 classifier to $\sim 12.5\%$ for the optimized RB2 classifier. It also shows that most of the improvements stem from the fact that we simply have a much larger training set as the survey has already been running for several years and because we have spectroscopically confirmed many of the sources. An additional improvement then comes from selecting the optimal set of features to avoid over-fitting to the training data and to remove highly correlated features.

\section{Application of RB2 to PTF data}\label{s:application}

  After going through the various steps of building the classification methodology in the previous sections, we now want to evaluate and present the performance of the real-bogus classifier. We start by fixing a training set of $50000$ sources, holding out $28448$ sources for the validation set.  In the following subsections, we will perform various tests and experiments of RB2 in the real-world use-case of discovering transient and variable events in PTF.
  
  \subsection{Classification performance}\label{ss:performance}
    
\begin{figure}
      \centering
    \includegraphics[width=\columnwidth,angle=0]{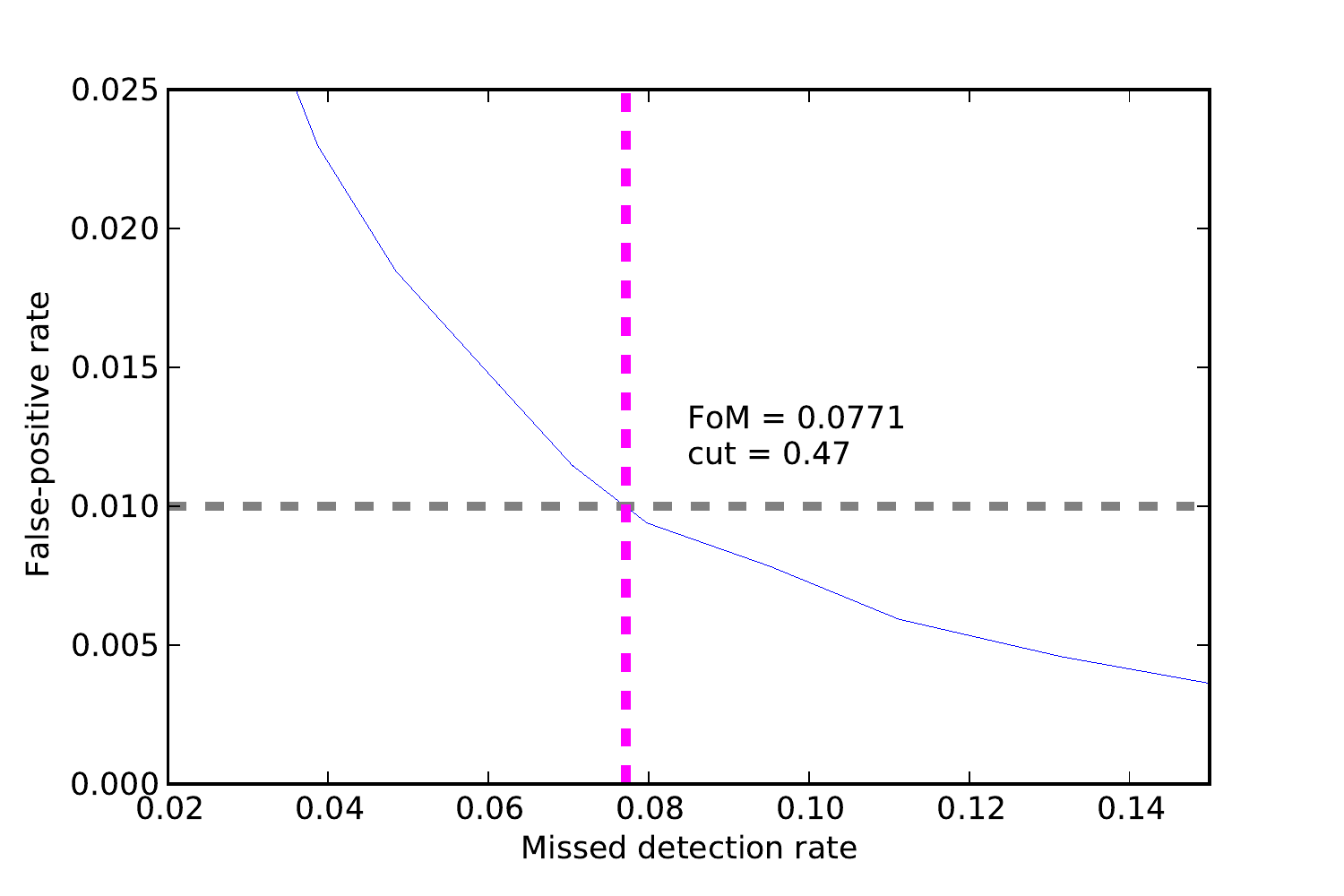}
     \caption{The ROC curve of the RB2 classifier applied to a fixed test set of 28448 PTF candidates. At a $1\%$ false-positive rate, we find a $p(B)$ threshold of $0.47$ and a missed-detection rate of $7.71\%$.}
      \label{f:performance_roc}
    \end{figure}

    The first order of business is to present the classification performance of the classifier trained with the new training set and applied to the new test set. Figure \ref{f:performance_roc} shows the now-familiar ROC curve, yielding a $7.71\%$ MDR at $1\%$ FPR at the probability threshold of $\tau = 0.53$. This is an improvement over the results obtained in the previous section due to the larger training set. While building the methodology of the framework, we have used smaller 5-fold cross-validated training sets, whereby our effective training sets were held to 80\% of their present size.  For the rest of the paper, we employ all 50000 training sources to unleash the full capacity of the real--bogus classifier in demonstrating actual performance in  realistic scenarios with PTF data. 
        
\begin{figure}
      \centering
    \includegraphics[width=\columnwidth,angle=0]{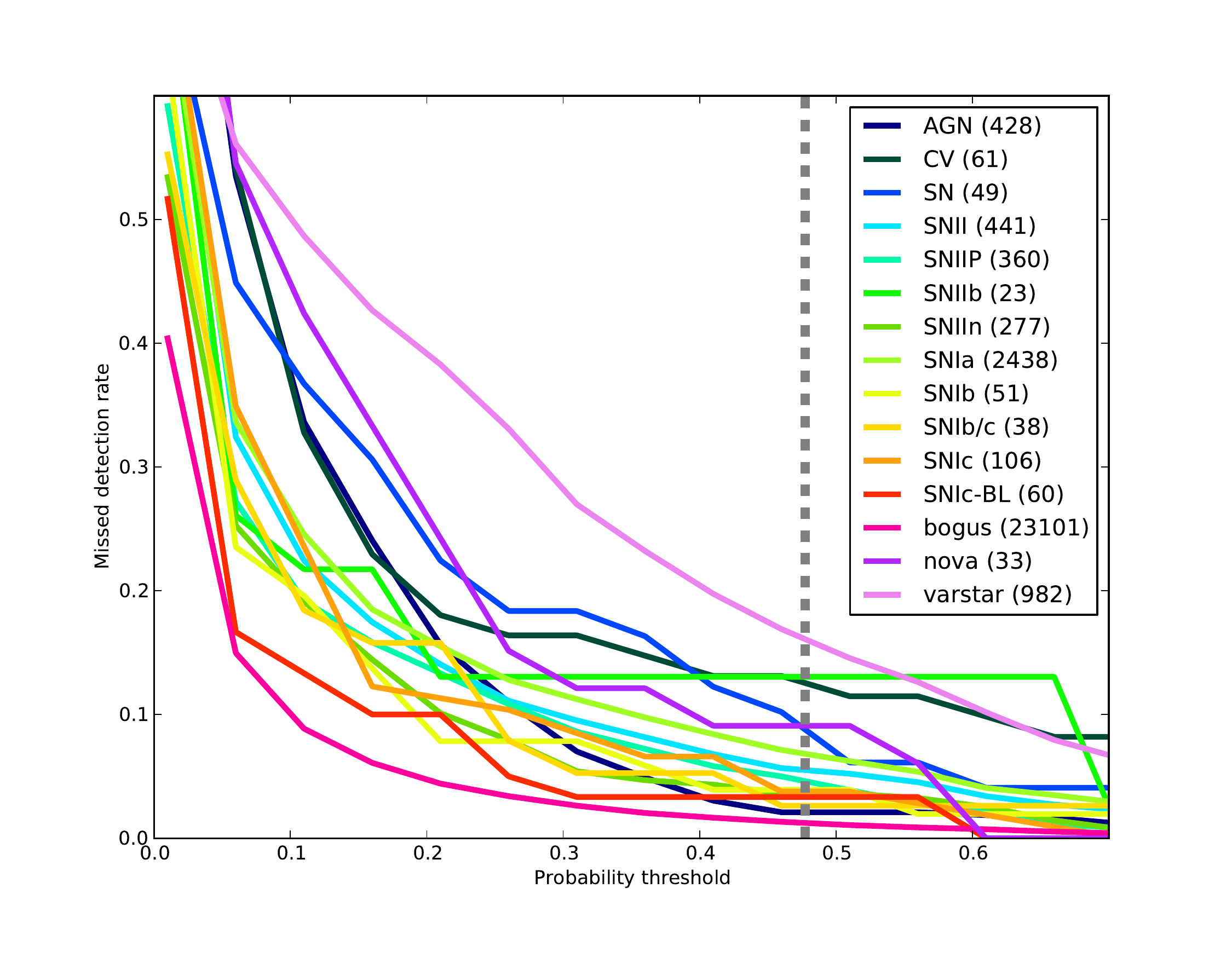}
     \caption{Performance, by object type, for 15 different types of object in PTF, including variable stars, novae, many kinds of supernovae, cataclysmic variables (CV) and active galactic nuclei (AGN) along with performance on bogus sources. The dashed vertical line marks the previously determined probability threshold that on average guarantees a 99\% purity of real classification across the entire data set while only missing $\sim 7.7\%$ of all real objects. This guarantee is, however, not valid for all subclasses of candidates and this figure gives an overview of the relative missed detection rate ranging from a few percent for bogus (true negatives) to $\sim 15\%$ for variable stars.}
      \label{f:classes_mdr}
    \end{figure}

    The mean performance of the classifier on the real--bogus problem is a 7.71\% MDR at 1\% FPR, but there are many kinds of transient and variable objects in PTF and we would like to get an impression of how the classifier handles each of these classes individually. From the description of the training set building in section \S\ref{ss:trainset}, it is clear that many real objects are supernovae, while there is a smaller fraction of variable stars and other types of variables, such as AGNs and CVs. Figure \ref{f:classes_mdr} shows the missed detection rate of all of these types as a function of probability threshold. Cutting at $p(B) \sim 0.47$, we obtain MDRs ranging from a few percent for various types of supernovae an AGN up to and MDR of $\sim 15\%$ for variable stars.  This poorer performance for variable stars is likely due to the fact that variable stars are typically observed much closer to their median reference brightness than transients, causing the detections to be close to the detection limit and thus more difficult to separate from bogus detections.

  \subsection{Deeper and earlier discovery}\label{ss:deeper}
  
    An important aspect in the discovery of transient events is how early the candidate can be detected and how soon potentially crucial follow-up resources can be deployed given the certainty of detection. This is complicated even more by the fact that rare (and possibly interesting) events are often either very nearby or very far away. By having included in the real--bogus training set all detections forwards and backwards in time around a confirmed candidate, we sacrifice some overall classification performance in order to improve the chance of detecting these rare events. In our validations, it is therefore useful to get a handle on how the detection pipeline performs as a function of candidate magnitude.
    
\begin{figure}
      \centering
    \includegraphics[width=\columnwidth,angle=0]{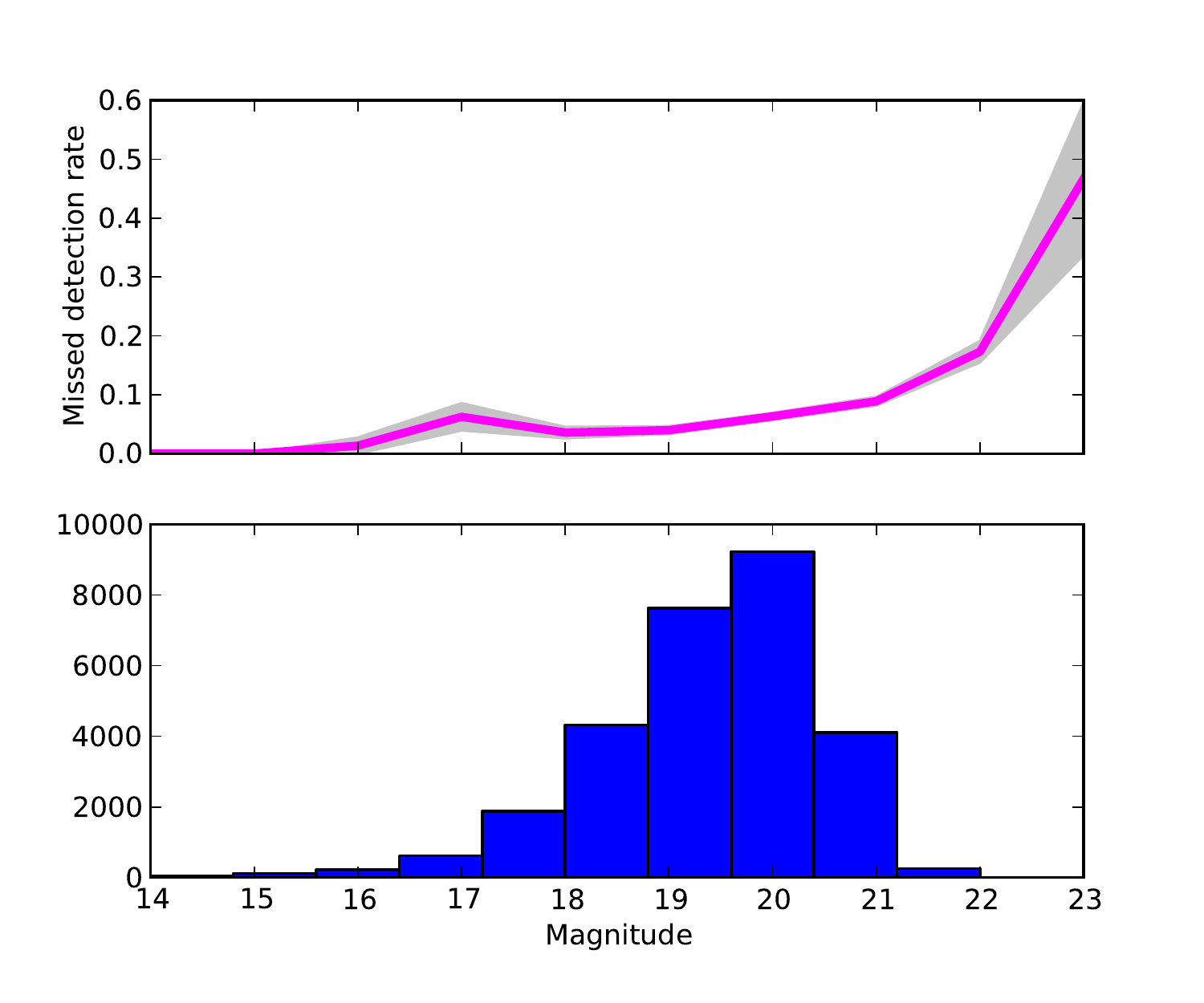}
     \caption{The magnitude histogram and the evolution of the missed detection rate with increasing magnitude (decreasing brightness). Going fainter increases the missed detection rate as expected. At the detector limit around magnitude 21, we expect to miss 10\% of the real candidates, quickly increasing along with the uncertainty. The shaded region shows the $\pm \sigma$ around the missed detection curve.}
      \label{f:mag_mdr}
    \end{figure}

    We first divide the validation set sources into 10 magnitude bins and limit the magnitude range based on the histogram of counts. We then calculate the missed detection rate for each of these bins at the optimal threshold, determined in section \S\ref{s:optim}, to mimic the selection process that these candidates would have gone through had this classifier been deployed at the time of their detection. The result is plotted in figure \ref{f:mag_mdr} where we see a $< 10\%$ risk of missing candidates until we get closer to the detection limit around magnitude $\gtrsim 21$, where the missed detection rate rises above 10\%.

\begin{figure}
      \centering
    \includegraphics[width=\columnwidth,angle=0]{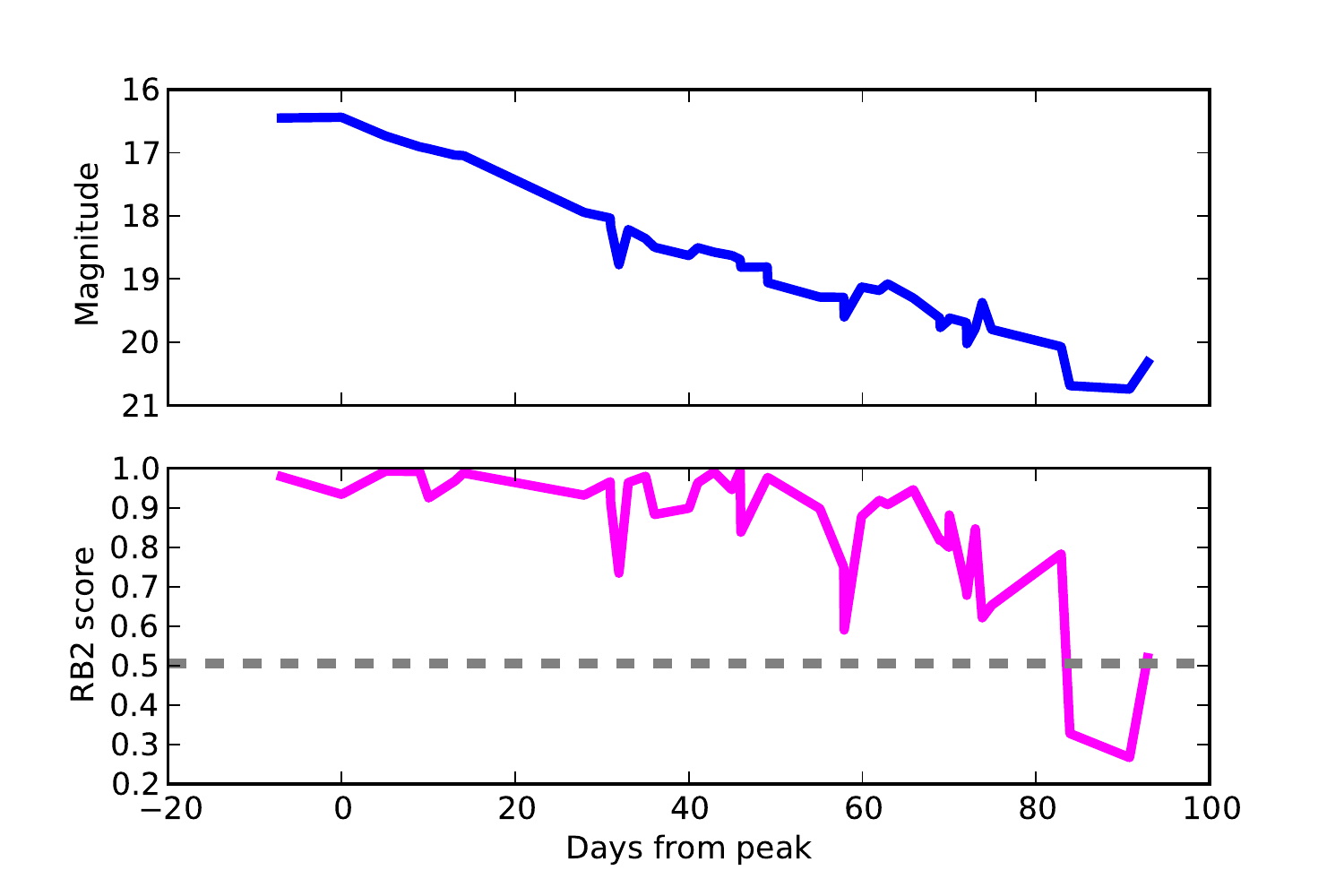}
     \caption{Evolution of magnitude and RB2 score with detections. The light-curve follows a typical path for a type Ia supernova. There is some scatter in the RB2 score, but we are consistently above the threshold for discovery (horizontal lines) until the faintest parts at the end. The dotted line shows the threshold $\tau$ above which candidates are considered for discovery.}
      \label{f:sn_mag_rb2}
    \end{figure}

    Inspection of the detection history of individual sources can demonstrate how our classifier would have performed had it been deployed in the live pipeline. To give an example of this we select a spectroscopically confirmed type Ia supernova from the validation set which includes a full light-curve with 42 total detections, and we plot in figure \ref{f:sn_mag_rb2} the evolution of the source brightness in magnitudes as well as the real--bogus classification score for each of the individual detections (using no knowledge of any previous or future detections).

\begin{figure}
      \centering
    \includegraphics[width=\columnwidth,angle=0]{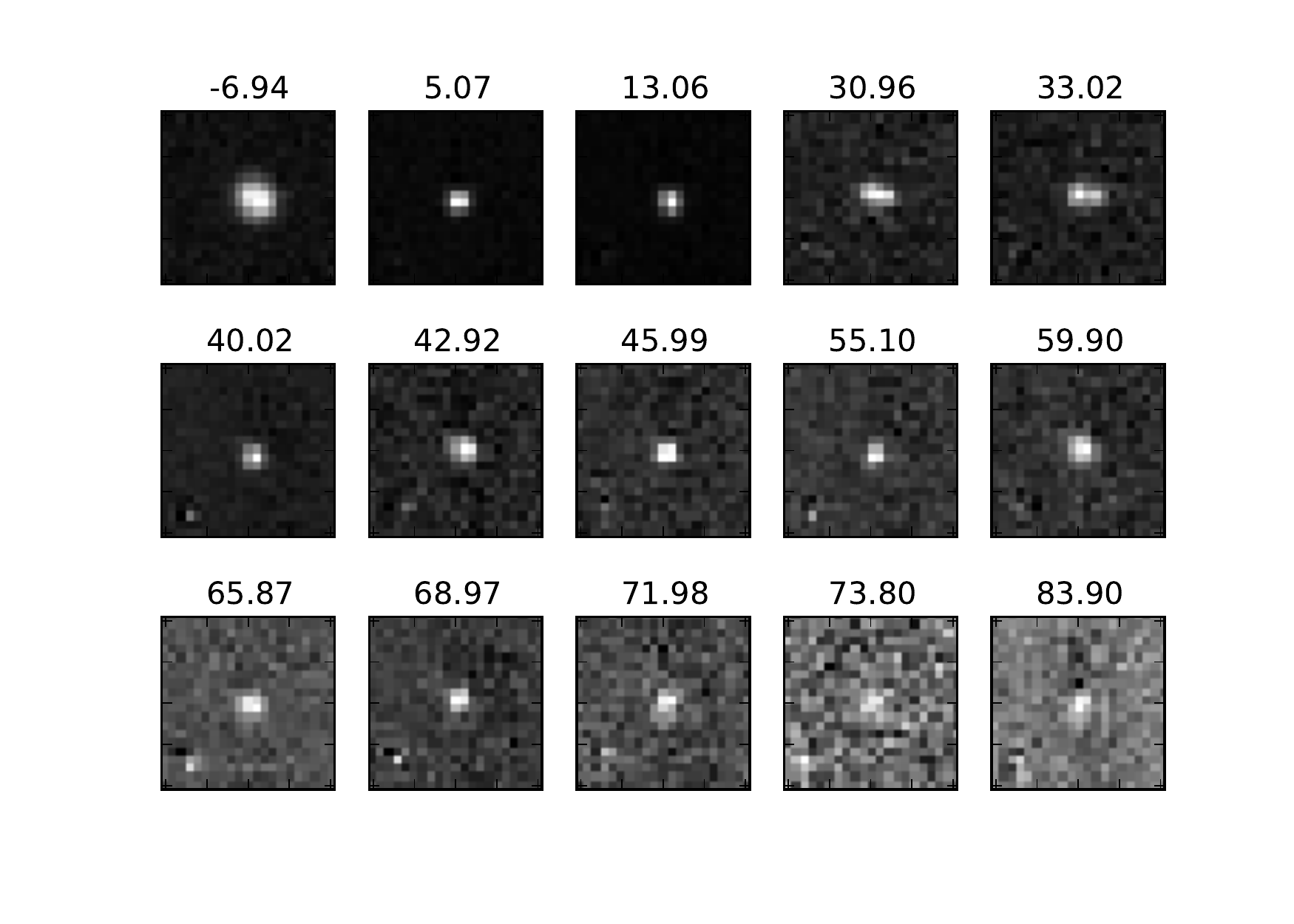}
     \caption{A subset of the stamps for the particular type Ia supernova discussed in the text. At late times the source is disappearing in low signal-to-noise conditions.}
      \label{f:sn_stamp}
    \end{figure}

    The real--bogus score is consistently above the threshold for classifying these candidates as real until the  last few detections of the supernova when the source disappears into the background at magnitude $\sim 21$. For inspection, we plot in figure \ref{f:sn_stamp} a subset of the 42 images for this particular supernova.
  
  \subsection{Application to 2011/12 detections}

    This framework is built to advance the way in which transient and variable events are discovered in large synoptic surveys where the data streams are quickly becoming more plentiful than any group of humans can handle. It is meant to be used in the real-time loop, ranking interesting candidates for inspection and efficient follow-up decisions. The training set that was built for this classification task was restricted to PTF 2010 data, so we are able to get a realistic measure of the performance by validating the classifier on 2011 and 2012 discoveries.

    In a realistic usage scenario, we will rely on the optimal threshold value determined in \S\ref{ss:performance}. In the previous section we claim that we should only miss $\sim 8\%$ reals below this threshold, with a trade-off contamination of $1\%$ bogus misclassified as reals. As with the initial training set, there is the unavoidable concern that the population of 2011/12 reals are biased by the previous incarnation (RB1) that missed actual reals. To keep this part simple, we will focus only the 2011/2012 supernovae discovered by PTF, so the question becomes how efficient would we have been at discovering these supernovae had this classifier been deployed at the time.
    
\begin{figure}
      \centering
    \includegraphics[width=\columnwidth,angle=0]{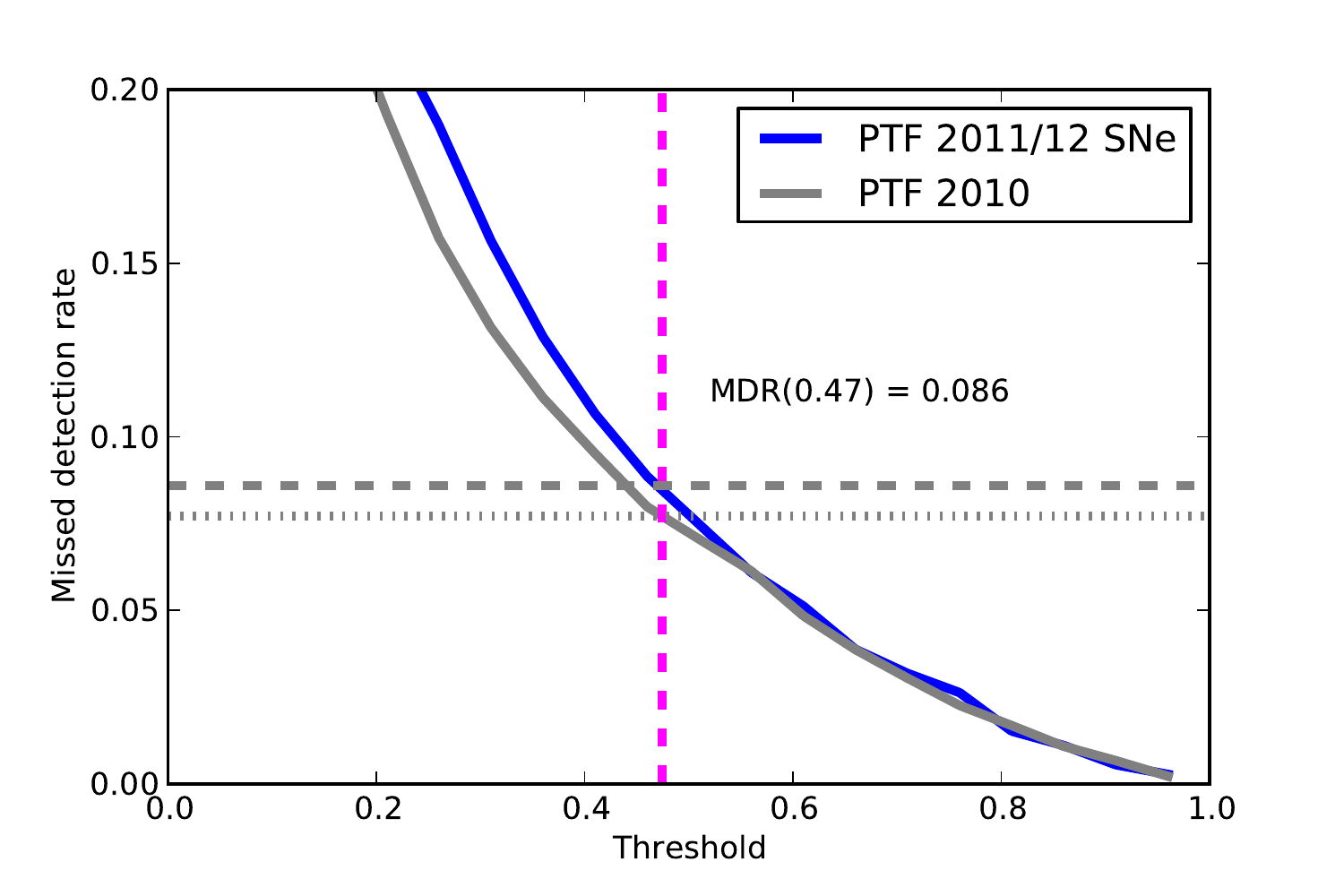}
     \caption{The missed detection rate of confirmed 2011/2012 supernovae as a function of decision threshold. At the determined threshold of 0.47, corresponding to a false-positive rate of 0.01 in the training set, we are close to the test-set missed detection rate of 7.7\% with a rate of 8.6\% for this validation set. This confirms the efficiency of the classifier in a realistic use-case for synoptic surveys where the threshold is pre-determined in the training phase and used to select candidates in incoming data.}
      \label{f:mds_validation}
    \end{figure}

    Figure \ref{f:mds_validation} shows the resulting missed detections as a function of the threshold parameter compared to the 2010 test. We do not build a set of bogus candidates for this validation, and therefore have no measure of the false-positive rate, as the optimal threshold has been determined in the training phase and we rely on this for a realistic discovery scenario. We note that the missed detection rate of $8.6\%$ at the previously determined probability threshold of $0.47$ is close to the $7.71\%$ expected from the analysis in section \S\ref{ss:performance}.

  \subsection{The real-to-bogus ratio in PTF}\label{ss:rbratio}

    With the real-bogus classifier in hand, we can now go on to estimate the real-to-bogus ratio of the PTF survey. In a synoptic survey, getting a handle on the amount of bogus produced by the experiment can be very important as it quantifies the requirements for the data analysis pipeline. In PTF, there is on the order of $10^6$ potential candidates hitting the pipeline every night, and in future surveys (such as LSST) there will be orders of magnitude more. With a framework like this we can get insight into the actual distribution of the number of real and bogus candidates and estimate how this is affected by the particular choice of figure of merit. The likely use-case of a framework like this is to produce a ranked list of candidates every night for human or machine follow-up, and minimizing the amount of junk while maximizing the scientific gain.

\begin{figure}
      \centering
    \includegraphics[width=\columnwidth,angle=0]{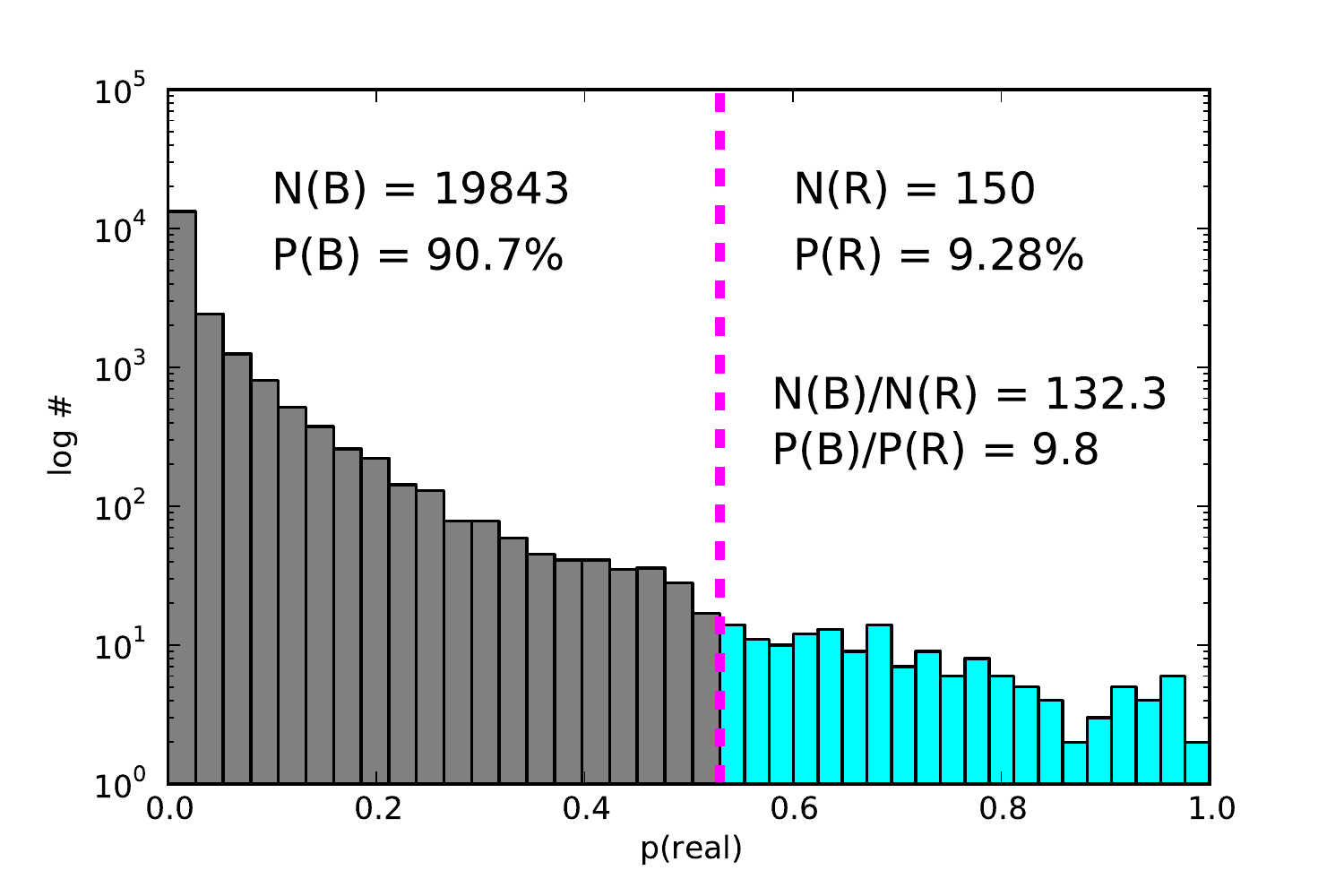}
     \caption{The distribution of predictions on a random sample of 20,000 candidates from 2011/12. We get a handle of the number of real to bogus candidates as determined by the RB2 classifier using the previously (\S\ref{ss:performance}) determined threshold of $\tau = 0.53$ (vertical line). We estimate that about 1 in 132 observations will be classified as real, corresponding to 1 in 10 when taking into account the figure of merit of 8.6\% MDR at 1\% FPR.}
      \label{f:r2b_ratio}
    \end{figure}

    We select a random set of $N=20,000$ sources from the PTF database from 2011-12. Figure \ref{f:r2b_ratio} shows the distribution of predictions by running these sources through the feature generation and RB2 prediction pipeline. By using the threshold determined in \S\ref{ss:performance} ($\tau = 0.53$), we find that $N(R)=150$ are classified as real and $N(B)=19843$ are classified as bogus\footnote{Stamps were unavailable for the remaining 7 sources.}, meaning that about 1 in 132 detections will be real. We also want to estimate the actual probability of encountering a real or bogus source in the discovery process, effectively weighting these numbers by the FPR and MDRs determined in section \S\ref{ss:performance}. We can estimate this using the Law of Total Probability, which states that the marginal probability of some event $A$ is the weighted average of the conditional probabilities of the possible outcomes $X$ over all possibilities.

    In our case the outcome of the measurement is the output classification as either a real or bogus discovery, so $X \in \{R,B\}$ and the event $A \in \{R,B\}$ signifies whether an object is truly real or bogus, respectively. The weights correspond to $P(R|X=R) = \mathrm{1-FPR}$, $P(R|X=B) = \mathrm{MDR}$, $P(B|X=B) = \mathrm{1-MDR}$ and $P(B|X=R) = \mathrm{FPR}$. As $P(X=R) = N(R) / N$ and $P(X=B)=N(B)/N$, this gives us the expression:

    \begin{eqnarray*}
      P(R) &= &P(R | X = R) P(X=R) \\ 
      && + P(R |X = B) P(X=B) \\ 
      &= &[1-\mathrm{FPR}] \frac{N(R)}{N} + \mathrm{MDR} \frac{N(B)}{N} \; , \; \textrm{and} \\
      P(B) &=&\mathrm{FPR} \frac{N(R)}{N} + [1 - \mathrm{MDR}] \frac{N(B)}{N} .
    \end{eqnarray*}
 
    Directly substituting in the numbers FPR $=0.01$ and MDR $=0.077$ from \S\ref{ss:performance} yields $P(R)=0.093$ and $P(B)=0.907$, giving a real-to-bogus ratio of $\sim$ 1 to 10.  However, as the FPR and MDR were estimated from the training data, these numbers may not be directly extendible to random samples of PTF data.  Consider the expression $P(R | X=R)\equiv \mathrm{1-FPR}$ used in the computation of $P(R)$.  In reality, we have estimated that on the training data, $P_{\rm train}(R | X=R) = 0.99$.  Rewriting this using Bayes' Theorem shows that
        \begin{eqnarray}
        \label{eqn:bayes_fpr}
P_{\rm train}(R | X=R) &=& \frac{P_{\rm train}(X=R | R) \, P_{\rm train}(R)}{P_{\rm train}(X=R)}.
    \end{eqnarray}
    First, since the training set was constructed with a larger proportion of reals (4.3-to-1 bogus-to-real ratio) than expected in PTF, it is clear that $P(R) < P_{\rm train}(R)$.  Second, since the training set of reals consisted primarily of observations of spectroscopically confirmed supernovae, and we find that supernovae are considerably easier to discover than variable stars (see figure \ref{f:classes_mdr}), we can safely assume that $P(X=R|R) < P_{\rm train}(X=R | R)$.  Thus, both terms in the numerator of (\ref{eqn:bayes_fpr}) should be smaller in the general population than for the training set.

What about the denominator?  Obviously, $P(X=R)$ will also be smaller than $P_{\rm train}(X=R)$, but by how much?  It is useful to rewrite the denominator of  (\ref{eqn:bayes_fpr}) using the Law of Total Probability, into
        \begin{eqnarray*}
  P_{\rm train}(X=R) &=& P_{\rm train}(X=R | R) \, P_{\rm train}(R)  + \\ &&P_{\rm train}(X=R | B) \, P_{\rm train}(B)
    \end{eqnarray*}
where the first part of the expression has already been analysed above.  For the second part, it is obvious that $P(B) > P_{\rm train}(B)$ using the same argument used above for the reals.  The bogus training set was attained by randomly selecting any detections that were not known to be real.   If all of these were in fact bogus, then $P_{\rm train}(X=R | B) \sim P(X=R | B)$; however, we know that this procedure causes some label noise, which would cause $P(X=R | B) \lesssim P_{\rm train}(X=R | B)$.
  
  Putting this all together, we deduce that it is likely that $P(R | X=R) < P_{\rm train}(R | X=R)$.  Similar analysis on $P(R | X=B)$ shows that it is also likely that $P(R | X=B) < P_{\rm train}(R | X=B)$, meaning that $P(R)$ is likely over-estimated.  Hence, we declare that the real-to-bogus ratio of 1:10 deduced by directly plugging in the FPR and MDR from the training set is an \emph{upper bound} on the true real-to-bogus ratio in PTF.  The unknown magnitudes of the effects of the various sample-selection biases in the training set preclude us from attaining a more precise estimate.  Only through more representative training samples may we attain better estimates of this ratio.
    
\section{Label contamination}
\label{s:contamination}

  We expect that some labels in our training set are wrong: a few candidates that are actually real are labelled as bogus and vice versa. The bogus sample was built by randomly selecting objects from the database that were not known reals, but there certainly exist reals that were missed by the previous pipeline, since we know that the MDR of the previous real--bogus classifier was non-negligible. Likewise, when constructing the real sample, for each confirmed  real we selected all candidates backwards and forwards in time  at the same spatial location (within 3.6 arcseconds).  In this process, we risk including a small number of bogus detections in the real sample due to false detections that occurred at that precise spatial location. The building of the training set is described in more detail in section \S\ref{ss:trainset}.

  Indeed, examining the top outliers in our classification---either sources labelled as real with highest classifier probability of bogus, or vice versa---reveals that some of these are obviously mislabelled.  Based on the manual examination of a few hundred of these, we crudely estimate that the purity of the sample is of order 99\%. This means that without employing any re-labeling or label-cleaning method we cannot expect any classifier to perform better than about 0.01 missed-detection rate at that same false-positive rate. Since the data encompass $78488$ thumbnails, it is impractical to weed out the mislabeled candidates by manually scanning all the objects. 
  
  Since label noise is a common problem in real-life machine learning problems,  in this section we explore the consequences of label contamination in both our real--bogus training and validation sets. The random forest classifier is believed to be relatively immune to training set contamination \citep{Brei01}, but we now investigate this in the context of our specific problem with our particular choice of figure of merit. 

  In order to quantify the effects of imperfect labelling on real--bogus classification performance, we run the following experiment.  For each of 20 levels of contamination, $\rho$, between 0.1\% to 15\%, we artificially contaminate the labels of the training or testing data by flipping the labels (real to bogus and vice versa) of a randomly-chosen subset of $\rho$ proportion of the data. To quantify the effect of this label contamination, we calculate the 5-fold cross-validation figure of merit for each of the following cases:
  \begin{enumerate}
\item On each fold, a random $\rho$ proportion of the \emph{training data} have their labels swapped.  The classifier is then fitted to those contaminated training data and evaluated on the unchanged left-out data.
\item On each fold, a random $\rho$ proportion of the left-out \emph{testing data} have their labels swapped.  The classifier is fitted to the unchanged training data and evaluated on the contaminated testing data.
\end{enumerate}
  
\begin{figure}\centering
    \includegraphics[width=0.9\columnwidth]{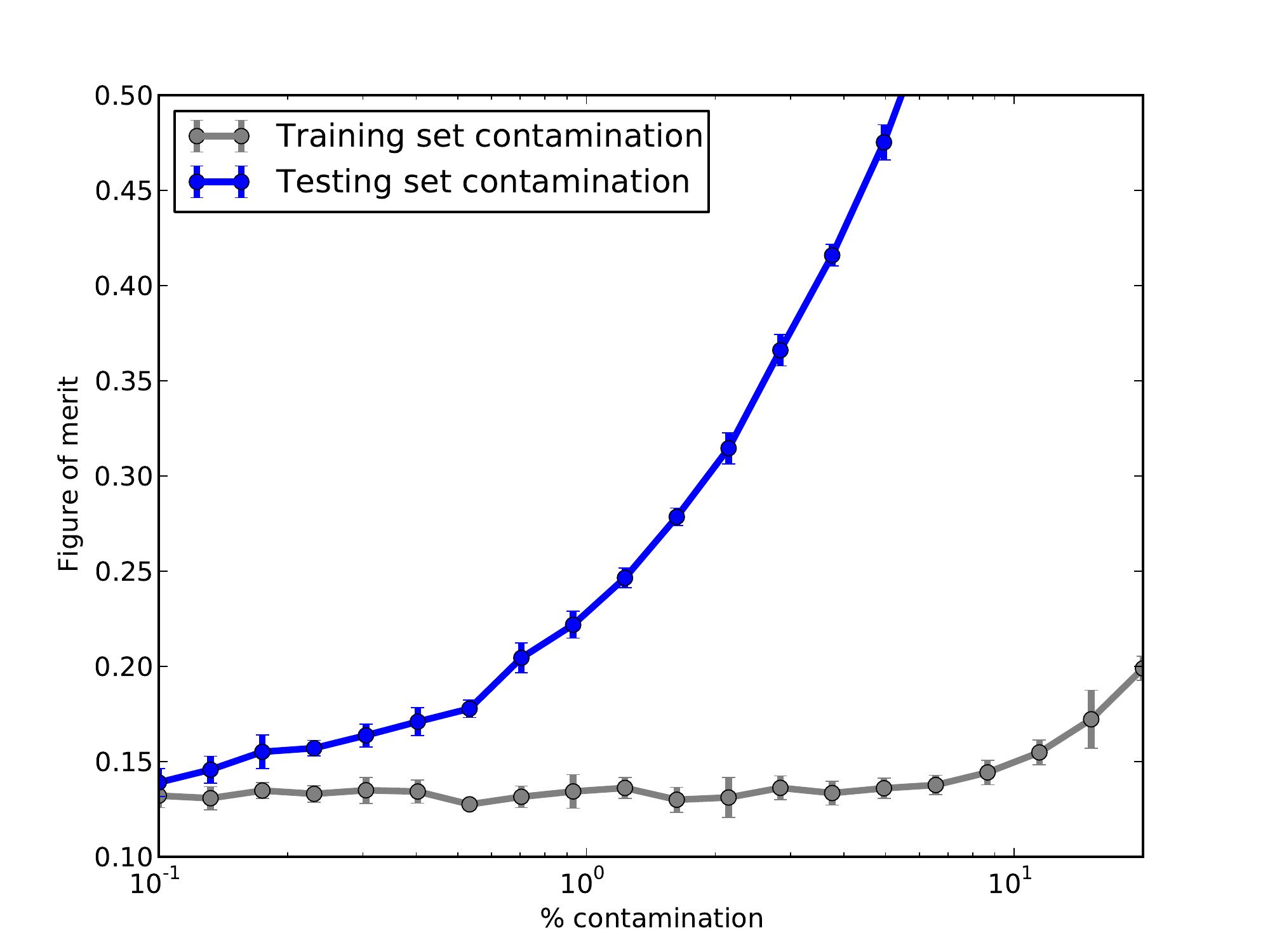}
    \caption{Artificial contamination of the dataset. The grey line shows the effect on the 5-fold cross-validated FoM from randomly flipping labels in the training set. We see that the classifier is relatively immune to dirty labels to almost 10\% contamination, and consistent with the FoM for the original training set, seen in figure \ref{f:feature_selection}. Contamination of the testing set (blue line) adversely affects the figure of merit, so a decrease in \emph{measured} model performance is expected if there is significant label contamination in the testing or validation set.}
    \label{f:contamination}
  \end{figure}

  In figure \ref{f:contamination}, we plot the 5-fold cross-validation figure of merit for each contamination proportion of the training data (grey) and testing data (blue).   The classifier is indeed immune to dirty labels in the training set up to a contamination of $\sim 10\%$, as the FoM does not deviate significantly from the base level of 0.13 (consistent with the 5-fold cross-validated FoM in figure \ref{f:feature_selection}).  This is an important insight for future surveys, as it allows one to bootstrap an event-discovery classifier with an imperfect training set in order to maximize the scientific output from the outset. We have demonstrated that it is more important to have a large, robust training set with some label noise than a small and limited yet perfectly labelled training set.  We recommend that future surveys take an inclusive approach to populating their training set of real and bogus detections.
  
\section{Conclusions}\label{s:conclusion}

    In this paper we have motivated the need for automated pipelines for the discovery of variable and transient phenomena in the synoptic survey era, where the data streams will be too immense for any group of humans to sift through. We have shown how this pipeline can be implemented using modern non-linear machine learning methods, enabling ranking of promising discoveries and subsequent deployment of expensive follow-up resources in real-time.

  Before we discuss some extensions and future possibilities of this real--bogus classification framework, we will conclude with a brief summary of the steps taken in this work, serving as a high-level mini-guide for machine-learned discovery for current and future subtraction-based imaging surveys where real-time detection is a priority.

  \begin{enumerate}

    \item Build the training set, a list of sources with known labels. In section \S\ref{ss:trainset} we describe how we built the initial training set, and in section \S\ref{s:contamination} we show that this does not have to be perfectly clean in order to be used by the random forest classifier. In figure \ref{f:performance} we see that the size of the training set is important, so the more labelled instances available the better. 

    \item Calculate features for the training set. Some of the features outlined in section \S\ref{ss:features} might not be useful for other surveys, so some domain knowledge is required in this step. 

    \item Pick a non-linear supervised classification scheme and train the classifier, holding out test and validation sets depending on the amount of sources available. In this project we have used a random forest classifier as outlined in section \S\ref{ss:method}, and shown that it outperforms a few other ML classifiers for this problem.  In general, a large training set and good features are more important than choosing a particular classifier, and there are a number of classifiers that one may consider.
    
    \item Use cross-validation to determine the optimal tuning parameters of the classifier. To do this, a particular figure of merit needs to be chosen that makes sense for the survey. In this paper we have used the missed detection rate at a false-positive rate of 1\%, because we want to avoid being swamped by false-positives, i.e., predicted real sources that are in fact bogus. 

    \item Select an optimal subset of features. We show in section \S\ref{ss:featselect} that this step can be crucial in order to squeeze the highest accuracy and generalizability out of the classifier. Simply using random forest feature importance is not sufficient for our problem, as this method does not take into account the correlation between features. We have used the iterative backwards selection method to select 23 out of the 42 features and gain almost 5\% in classification performance on the test and validation sets.

    \item When new observations arrive, a set of features is calculated for the given source and fed to the classifier for prediction. The classifier will output a probability of the source being real or bogus that can be used for ranking for further follow-up by other automated software of humans. Using the previously obtained figures we expect 99\% of the sources above $\tau=0.53$ to be real, while missing around 7.7\% that did not make this threshold.

  \end{enumerate}

  No machine learner can do well on uninformative data, and beyond some obvious choices any classifier will only be as good as the features fed to it. Coming up with useful features is a challenge that often require a lot of domain knowledge, and in the field of astronomy, often a good deal of experience with image processing. This project is of course no exception, and the performance of the classifier can no doubt be improved upon with more and better (more is not always better, see \S\ref{ss:featselect}) features. In particular, there has been many developments in the field of computer vision that share many of the same characteristics with problems in astronomy \citep{lowe2004,dalal2005}, and we feel that applying this knowledge would be an obvious way to improve on future applications of machine learning and automation in astronomical imaging.

  In PTF, there is an additional component that can be used for discovery that we touched upon while building the training set in section \S\ref{ss:trainset}. The telescope will attempt to return to the same part of the sky twice every night, in an attempt to discover asteroids. This can be a help in the real-bogus step, as the likelihood of a real source disappearing within a couple of hours is small. In this paper we have focused on the discovery of single detections in a real-time setup, and not the classification of objects from what eventually becomes full lightcurves. This is in itself an active area of research, see e.g. \citet{richards2012macc}.

  Another prospect of this framework is to further investigate the possibility of using existing training data and classifiers in the start-up phase of new surveys. We touched upon this in section \S\ref{ss:trainset} and section \S\ref{s:contamination}, where we showed that the RB2 classifier is relatively immune to dirty labels up to a contamination at the 10\% level. This means that it is possible to build an initial training set that is not perfect, but has a larger number of instances (more training data is better, see \S\ref{ss:evaluation}). In the world of machine learning, several exciting developments might enable a more formal treatment of this issue. Recently, \citet{richards2012} showed that the field of \emph{Active Learning} (see \citet{settles2010} for recent review) can be effectively applied against the sample selection bias to select only a small number of examples from the new instrument as to maximally improve the classifier by labeling and augmenting the ``old'' training set with these samples. 
  The related field of \emph{Transfer Learning} \citep{pan2010} deals with changes in the feature distributions across machine learners. More research in this area and the application of to problems in astronomy might offer an exciting opportunity for new surveys to solve the cold-start problem in more automated ways than is currently possible.

Lastly, many of the bogus detections arise because of bad image subtraction, so advances in this area would lower the amount of bogus detections. For large-scale future surveys a better image subtraction method is obviously something that needs to be considered. Because of the noisy nature of the data and the need for intelligent deployment of follow-up resources, however, the need for statistically verified detection measures that can be used to rank a subset of the candidate events continues to be very important.

\section*{Acknowledgments}
The authors acknowledge the generous support of a CDI grant (\#0941742) from the National Science Foundation.
The Palomar Transient Factory project is a scientific collaboration between the California
Institute of Technology, Columbia University, Las Cumbres Observatory, the Lawrence Berkeley
National Laboratory, the National Energy Research Scientific Computing Center, the University of
Oxford, and the Weizmann Institute of Science.
The National Energy Research Scientific Computing Centers (NERSC), supported by the Office of Science of the U.S. Department of Energy, provided computational resources and data storage for this project.
This research has made use of the VizieR catalogue access tool, CDS, Strasbourg, France. 


\bibliography{bibfile}

\label{lastpage}

\end{document}